\documentclass[11pt]{article}
\usepackage[final]{acl}

\usepackage{times}
\usepackage{latexsym}
\usepackage[T1]{fontenc}
\usepackage[utf8]{inputenc}
\usepackage{microtype}
\usepackage{inconsolata}
\usepackage{graphicx}
\usepackage{booktabs}
\usepackage{amsmath,amssymb}
\usepackage{multirow}
\usepackage{xspace}
\usepackage{array}

\newcommand{\ours}{NanoVDR\xspace}
\newcommand{\ndcg}{NDCG@5\xspace}
\newcommand{\ret}[1]{$_{\text{\tiny(#1\%)}}$}

\title{\ours: Distilling a 2B Vision-Language Retriever into a 70M Text-Only Encoder for Visual Document Retrieval}

\author{Zhuchenyang Liu, Yao Zhang, Yu Xiao \\
  Aalto University \\
  Espoo, Finland \\
  \texttt{zhuchenyang.liu@aalto.fi}}

\begin{document}
\maketitle

\begin{abstract}
Vision-Language Model (VLM) based retrievers have advanced visual document retrieval (VDR) to impressive quality. They require the same multi-billion parameter encoder for both document indexing and query encoding, incurring high latency and GPU dependence even for plain-text queries.
We observe that this design is unnecessarily symmetric: documents are visually complex and demand strong visual understanding, whereas queries are just short text strings.
\ours exploits this query--document asymmetry by decoupling the two encoding paths: a frozen 2B VLM teacher indexes documents offline, while a distilled text-only student as small as 69M parameters encodes queries at inference.
The key design choice is the distillation objective.
Through systematic comparison of six objectives across three backbones and 22 ViDoRe benchmark datasets, we find that pointwise cosine alignment on query text consistently outperforms ranking-based and contrastive alternatives, while requiring only pre-cached teacher query embeddings and no document processing during training.
Furthermore, we identify cross-lingual transfer as the primary performance bottleneck, and resolve it cheaply by augmenting training data with machine-translated queries.
The resulting \ours-S-Multi (DistilBERT, 69M) retains 95.1\% of teacher quality and outperforms DSE-Qwen2 (2B) on v2 and v3 with 32$\times$ fewer parameters and 50$\times$ lower CPU query latency, at a total training cost under 13 GPU-hours.
\end{abstract}

\begin{figure*}[t]
    \centering
    \includegraphics[width=\textwidth]{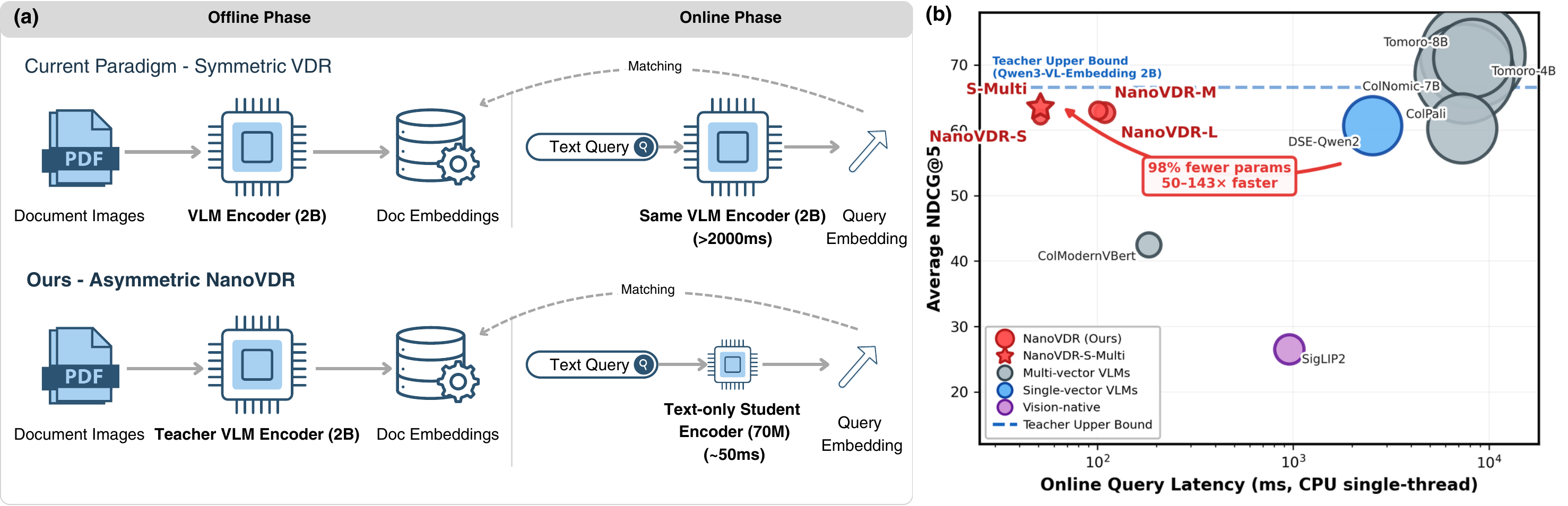}
    \caption{\textbf{Motivation and deployment advantage of \ours.}
    (a)~\textbf{Symmetric vs.\ asymmetric retrieval:} Current VDR systems (top) use the same heavy VLM encoder (2B) for both offline document indexing and online query encoding ({$>$}2{,}000\,ms per query). \ours (bottom) decouples the two: the frozen VLM teacher encodes documents offline, while a distilled text-only student (70M) encodes queries online in {$\sim$}50\,ms on CPU.
    (b)~\textbf{Performance vs.\ latency:} On the ViDoRe benchmark (mean \ndcg across v1/v2/v3), \ours models achieve near-teacher accuracy (dashed line) at 50--143$\times$ lower CPU latency. Bubble size is proportional to model parameter count; the star ($\bigstar$) marks \ours-S-Multi, the multilingual-augmented variant (\S\ref{sec:multilingual-augment}).}
    \label{fig:teaser}
\end{figure*}

\section{Introduction}
\label{sec:intro}

Visual document retrieval (VDR) has achieved remarkable effectiveness in retrieving information from visually rich documents---financial reports with charts, scientific papers with figures, industrial manuals with diagrams---by treating each page as an image rather than relying on OCR-based text extraction \cite{faysse2025colpaliefficientdocumentretrieval,ma2024unifyingmultimodalretrievaldocument}.
State-of-the-art systems use Vision-Language Models (VLMs) to encode both queries and document pages into a shared embedding space \cite{faysse2025colpaliefficientdocumentretrieval,ma2024unifyingmultimodalretrievaldocument,tomoro2025colqwen3,nomic2025colnomic}.

However, these systems apply the same heavyweight VLM encoder for both document indexing and query encoding.
This results in high computational overhead at query time, requiring multi-billion parameter models and GPU inference even for plain-text queries, and leads to large index storage costs for high-dimensional representations.

A key observation is that this design is unnecessarily symmetric: documents are visually complex and genuinely require strong visual understanding, whereas queries are short text strings that carry no visual content.
Using a multi-billion parameter VLM to encode text-only queries wastes the model's visual processing capacity entirely.

To exploit this query--document asymmetry, we propose \ours,\footnote{Code and training scripts: \url{https://github.com/Ryenhails/NanoVDR}. Model checkpoints and training data: \url{https://huggingface.co/nanovdr}. Interactive demo: \url{https://huggingface.co/spaces/nanovdr/NanoVDR-Demo}.} which decouples the two encoding paths through knowledge distillation (Figure~\ref{fig:teaser}a).
A frozen VLM teacher indexes documents offline, producing single-vector visual embeddings; a lightweight text-only student (as small as 69M parameters) encodes queries at inference by mapping them into the teacher's embedding space via a learned projector.
The student requires no vision module and runs on CPU in ${\sim}$50\,ms, enabling single-vector cosine similarity retrieval.

The central design choice is the distillation objective---how to train the student to faithfully represent queries in the teacher's visual space.
Through systematic comparison of six objectives across three backbones and the full 22-dataset ViDoRe benchmark, we make the following contributions:
\begin{itemize}
    \item \textbf{Asymmetric cross-modal distillation.} We propose a framework that distills a frozen 2B VLM teacher into text-only student encoders (69--151M) for VDR. Building on prior work showing that embedding alignment is a strong recipe for asymmetric query-encoder distillation within text retrieval \cite{kim2023embeddistillgeometricknowledgedistillation,wang-hong-2023-query,vujanic-ruckstiess-2026-leaf}, we show that the same objective transfers across modalities: pointwise cosine alignment, which directly matches student and teacher query embeddings, outperforms all ranking-based and contrastive objectives while requiring only pre-cached teacher query embeddings and eliminating corpus-related processing from training entirely.

    \item \textbf{Extreme efficiency.} \ours-S (DistilBERT, 69M) outperforms DSE-Qwen2 (2B) on v2 and v3 with 32$\times$ fewer parameters and 50$\times$ lower query latency (Figure~\ref{fig:teaser}b), at a total training cost under 13 GPU-hours. It also keeps the storage efficiency compared to multi-vector methods, which is inherited from the teacher model.
    \item \textbf{Cross-lingual augmentation.} We further identify in asymmetric encoders for VDR task, the cross-lingual transfer is the primary performance bottleneck, rather than cross-modal transfer. We resolve it via query-only multilingual augmentation, raising teacher retention from 92.4\% (\ours-S) to 95.1\% (\ours-S-Multi) at low cost.
    
\end{itemize}

\begin{figure*}[ht]
    \centering
    \includegraphics[width=\textwidth]{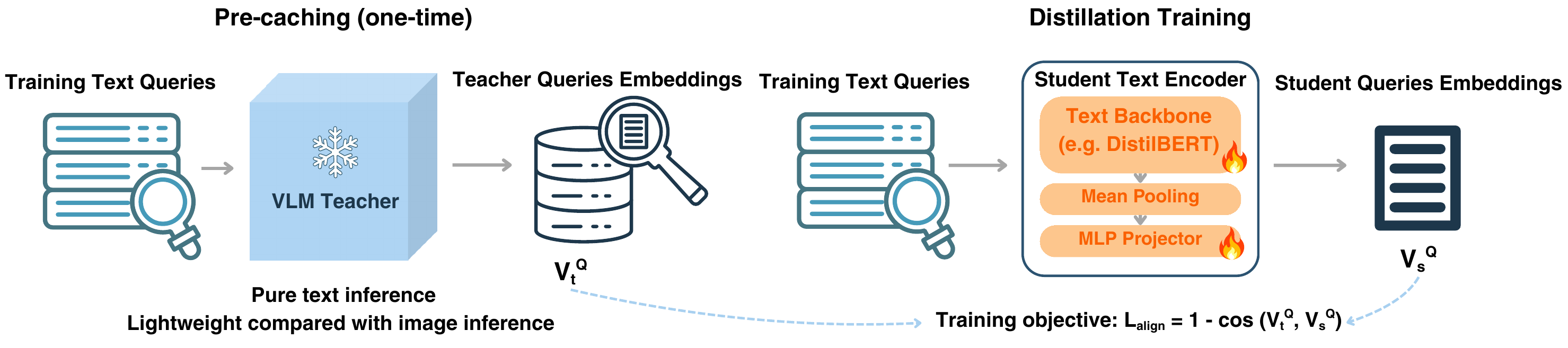}
    \caption{\textbf{Query-centric distillation training of \ours.}
    \textbf{Left:} The frozen VLM teacher pre-caches training query embeddings via text-only inference.
    \textbf{Right:} The student text encoder is trained to minimize $\mathcal{L}_\text{align} = 1 - \cos(\mathbf{v}^Q_t, \mathbf{v}^Q_s)$, requiring no document images or negative sampling.}
    \label{fig:framework}
\end{figure*}

\section{Related Work}
\label{sec:related}

\subsection{Visual Document Retrieval}

Visual document retrieval treats document pages as images and uses VLMs for both query and document encoding.
ColPali \cite{faysse2025colpaliefficientdocumentretrieval} adapts PaliGemma into a ColBERT-style \cite{khattab2020colbertefficienteffectivepassage} late interaction model, where each document page produces hundreds of token-level embeddings scored via MaxSim.
The same work introduced the ViDoRe benchmark spanning diverse document types.
DSE \cite{ma2024unifyingmultimodalretrievaldocument} takes a single-vector approach, producing one embedding per document screenshot using Qwen2-VL \cite{wang2024qwen2vlenhancingvisionlanguagemodels}.
VisRAG \cite{yu2025visragvisionbasedretrievalaugmentedgeneration} demonstrates retrieval-augmented generation over visual documents.
While retrieval quality has steadily improved, each generation of models has grown larger: more recent multi-vector systems based on 4--8B VLMs \cite{tomoro2025colqwen3,nomic2025colnomic} achieve the highest quality but with query latency exceeding 7 seconds on CPU, further widening the efficiency gap.
Vision-native encoders such as SigLIP2 \cite{tschannen2025siglip2multilingualvisionlanguage} and Jina-CLIP \cite{koukounas2024jinaclipclipmodel} offer lighter alternatives but substantially lag behind VLM-based approaches on document retrieval tasks.

Three concurrent works attempt to bridge efficiency and visual understanding.
VISTA \cite{zhou-etal-2024-vista} augments a frozen text encoder with a ViT image tokenizer, enabling multi-modal retrieval without modifying the text backbone; however, the ViT remains required at inference, and the model has not been evaluated on document-level benchmarks.
ModernVBERT \cite{teiletche2025modernvbertsmallervisualdocument} builds a purpose-designed 250M vision-language encoder by fusing a SigLIP2 vision encoder with a ModernBERT backbone via early fusion, matching ColPali-level quality at 12$\times$ fewer parameters; nevertheless, both query and document encoding still require the full vision-language model.
SERVAL \cite{nguyen-etal-2025-serval} takes a generate-then-encode approach: a VLM generates textual descriptions of document images, which are then indexed by a standard text encoder. While zero-shot and effective, the pipeline requires massive VLM inference for every document at indexing time.

Another parallel line of work attacks the cost of multi-vector VDR from the document side, by shrinking what has to be stored per page: merging patch embeddings \cite{ma-etal-2025-towards-storage}, pruning them adaptively \cite{yan2025docprunerstorageefficientframeworkmultivector, liu2026structuralanchorpruningtrainingfree} or replacing them with a learned and resizable set of tokens \cite{xiao2026metaembedscalingmultimodalretrieval}.
These methods change what the index holds while leaving the retrieval architecture intact, so the full multi-billion-parameter VLM still runs at both document and query time.

Our approach differs: we distill the VLM's embedding space directly into a tiny text-only encoder, requiring neither a vision module at inference nor VLM-scale caption generation.

\subsection{Knowledge Distillation in Dense Retrieval}

Knowledge distillation \cite{hinton2015distillingknowledgeneuralnetwork} has been extensively applied to dense text retrieval, where using separate encoders for queries and documents is well-established (DPR \cite{karpukhin2020densepassageretrievalopendomain}, ColBERT \cite{khattab2020colbertefficienteffectivepassage}).
\ours extends this asymmetry across modalities, pairing a VLM document encoder with a text-only query encoder.
TAS-B \cite{hofstätter2021efficientlyteachingeffectivedense} uses topic-aware sampling with balanced training from a cross-encoder teacher.
MarginMSE \cite{hofstätter2021improvingefficientneuralranking} distills pairwise margin scores to train efficient bi-encoders.
RankDistil \cite{pmlr-v130-reddi21a} applies listwise KL-divergence with curriculum learning.
These approaches operate within a single modality (text-to-text) and rely on ranking-based objectives.
In the vision-language domain, CLIP-KD \cite{yang2024clipkdempiricalstudyclip} and TinyCLIP \cite{wu2023tinyclipclipdistillationaffinity} compress CLIP models via combinations of feature alignment and affinity mimicking, but target image classification rather than document retrieval.

Making the query side of a dense retriever cheap by aligning it to a frozen teacher's embedding space is an established recipe within text retrieval.
EmbedDistill \cite{kim2023embeddistillgeometricknowledgedistillation} distils the geometry of the teacher's embedding space rather than its scores; \citet{wang-hong-2023-query} show that aligning query embeddings alone already suffices to shrink the online encoder; and LEAF \cite{vujanic-ruckstiess-2026-leaf} trains a small student against a frozen embedding teacher without relevance judgments, hard negatives, or a contrastive term.
All three ask how to make the query encoder small, and all three answer within a single modality.

Most closely related is Unveil \cite{sun-etal-2025-unveil}, which distils a VLM teacher into a VLM student, so the student still encodes images. Our contribution is the cross-modal extension of the same recipe: the teacher is a vision-language model and the student is text-only, so the query encoder inherits the teacher's document embedding space without ever processing an image.

\section{Methodology}
\label{sec:method}

\subsection{System Overview}

Figure~\ref{fig:teaser}a illustrates the overall architecture.
Given a corpus of $N$ document pages, each rendered as an image $d_j$, and a text query $q$, visual document retrieval aims to rank pages $\mathcal{D} = \{d_1, \ldots, d_N\}$ by relevance to $q$.
\ours decouples the two encoding paths entirely:
a frozen VLM teacher $g$ indexes each page image offline as $\mathbf{v}^D_j = g(d_j) \in \mathbb{R}^d$,
while a lightweight text-only student $f_\theta$ encodes queries online as $\mathbf{v}^Q_s = f_\theta(q) \in \mathbb{R}^d$.
Retrieval is performed via cosine similarity: $\text{score}(q, d_j) = {\mathbf{v}^Q_s}^\top \mathbf{v}^D_j$.

Following Sentence-Bert~\cite{reimers2019sentencebertsentenceembeddingsusing}, the student text encoder consists of a pre-trained backbone $h$, mean pooling, and a two-layer MLP projector:
\begin{equation}
    f_\theta(q) = \text{norm}\!\big(\text{MLP}(\text{pool}(h(q)))\big)
\end{equation}
where $\text{MLP}(\mathbf{x}) = W_2 \, \sigma(W_1 \mathbf{x} + b_1) + b_2$ with GELU activation.
The teacher remains completely frozen throughout; specific model choices are detailed in \S\ref{sec:setup}.

\subsection{Query-Centric Distillation}
\label{sec:align}

Figure~\ref{fig:framework} illustrates the training pipeline, which proceeds in two stages (left to right).
First, the frozen VLM teacher encodes all training queries in text-only mode, producing target embeddings $\mathbf{v}^Q_t = g(q) \in \mathbb{R}^d$. Second, the student text encoder is trained to produce query embeddings close to the teacher's.
The alignment loss directly minimizes the angular distance:
\begin{equation}
    \mathcal{L}_\text{align} = 1 - \frac{\mathbf{v}^Q_s \cdot \mathbf{v}^Q_t}{\|\mathbf{v}^Q_s\| \, \|\mathbf{v}^Q_t\|}
\end{equation}
Because the teacher maps both queries and documents into the same embedding space, training the student to match teacher query embeddings simultaneously enables retrieval against teacher document embeddings, despite the student never seeing any images.
This pointwise formulation requires no document embeddings, no negative sampling, and no corpus-level processing.

A key practical advantage of alignment-only distillation ($\mathcal{L}_\text{align}$) is that it requires only teacher query embeddings, which are text-encoded.
Ranking-based objectives additionally require teacher document embeddings ($\mathcal{L}_\text{rank}$, Eq.~\ref{eq:rank}) to construct in-batch similarity distributions, necessitating the teacher to process every training image---the dominant bottleneck in the pre-caching pipeline.

\subsection{Multilingual Query Augmentation}
\label{sec:multilingual-augment}

Because alignment training is purely query-centric, extending the student to new languages requires only additional query text---not new document images or teacher re-encoding.
We translate ${\sim}$489K English training queries into five target languages (Portuguese, Spanish, German, French, Italian) using Helsinki-NLP Opus-MT models \cite{tiedemann-thottingal-2020-opus}, balancing each language to ${\sim}$200K queries.
Each translated query is re-encoded by the frozen teacher in text mode, producing a new target embedding.
The augmented dataset combines these 778K translations with the original 711K pairs, yielding 1.49M training pairs (details in Appendix~\ref{app:multilingual-pipeline}).

\begin{table*}[t]
\centering
\small
\setlength{\tabcolsep}{3pt}
\caption{Main results (\ndcg $\times$100) on the ViDoRe benchmark. Each benchmark version is averaged over the indicated number of datasets. Params = total model parameters; for \ours this includes the backbone + MLP projector. Scoring = retrieval scoring method: MaxSim (token-level late interaction) or Cosine (single-vector dot product). $\dagger$: Qwen3-VL-Embedding-2B \cite{li2026qwen3vlembeddingqwen3vlrerankerunifiedframework}, our frozen teacher used for offline document indexing. Best per group in \textbf{bold}. Subscripts on \ours rows indicate teacher retention (\%, student/teacher). $\ddagger$: SERVAL \cite{nguyen-etal-2025-serval} generates a textual description of every page offline with a VLM and indexes those descriptions with a standard text encoder. Its ViDoRe v2 score is recomputed from the per-dataset results reported in that paper; see Appendix~\ref{app:serval}.}
\label{tab:main}
\begin{tabular}{llllrrr}
\toprule
Model & Backbone & Params & Scoring & ViDoRe v1 (10) & ViDoRe v2 (4) & ViDoRe v3 (8) \\
\midrule
\multicolumn{7}{l}{\textit{Multi-vector VLMs (MaxSim scoring, multi-token document representation)}} \\
\quad Tomoro-8B       & ColQwen3-8B      & 8.8B & MaxSim & \textbf{90.6} & 65.0          & \textbf{59.0} \\
\quad Tomoro-4B       & ColQwen3-4B      & 4.0B & MaxSim & 90.2          & \textbf{65.2} & 57.6          \\
\quad ColNomic-7B     & ColQwen2.5-7B    & 7.8B & MaxSim & 89.8          & 60.4          & 55.9          \\
\quad ColPali         & PaliGemma-3B     & 3.0B & MaxSim & 84.2          & 54.7          & 42.0          \\
\quad ColModernVBert  & ModernBERT-base  & 250M & MaxSim & 76.7          & 33.4          & 17.4          \\
\midrule
\multicolumn{7}{l}{\textit{Single-vector VLMs (Cosine scoring, one vector per document)}} \\
\quad DSE-Qwen2       & Qwen2-VL-2B      & 2.0B & Cosine & \textbf{85.1} & 55.7          & 41.3          \\
\quad Qwen3-VL-Emb$^\dagger$ & Qwen3-VL-2B & 2.0B & Cosine & 84.3       & \textbf{65.3} & \textbf{50.0} \\
\midrule
\multicolumn{7}{l}{\textit{Vision-native encoders (Cosine scoring, visual-text contrastive pre-training)}} \\
\quad JinaCLIP        & CLIP-ViT-L       & 400M & Cosine & \textbf{53.7} & \textbf{26.7} & \textbf{20.7} \\
\quad SigLIP2         & So400m           & 1.1B & Cosine & 44.6          & 20.1          & 14.8          \\
\quad BiModernVBert   & ModernBERT-base  & 250M & Cosine & 37.4          & 10.9          &  5.5          \\
\midrule
\multicolumn{7}{l}{\textit{Generate-then-encode (VLM caption $\to$ text encoder)}} \\
\quad SERVAL$^\ddagger$ & Qwen2.5-VL-32B + INF-7B & 32B+7B & Cosine & --- & 62.2$^\ddagger$ & --- \\
\midrule
\multicolumn{7}{l}{\textit{\ours (text-only student, Cosine scoring, distilled from Qwen3-VL-Emb$^\dagger$)}} \\
\quad \ours-L         & ModernBERT-base  & 151M & Cosine & \textbf{82.4}\ret{97.7} & 61.5\ret{94.2}          & 44.2\ret{88.4}          \\
\quad \ours-M         & BERT-base        & 112M & Cosine & 82.1\ret{97.4}          & \textbf{62.2}\ret{95.3} & 44.7\ret{89.4}          \\
\quad \ours-S         & DistilBERT       &  69M & Cosine & 82.2\ret{97.5}          & 60.5\ret{92.6}          & 43.5\ret{87.0}          \\
\quad \ours-S-Multi & DistilBERT    &  69M & Cosine & 82.2\ret{97.5}          & 61.9\ret{94.8}          & \textbf{46.5}\ret{93.0} \\
\bottomrule
\end{tabular}
\end{table*}

\begin{table*}[t]
\centering
\small
\setlength{\tabcolsep}{4pt}
\caption{Deployment cost comparison (query-time). All latency measured on a single CPU thread (batch size 1). Params = total parameters. Dim = embedding dimension per vector. Tok/Doc = tokens stored per document page (1 = single-vector). Size = query encoder checkpoint (float32); for \ours this is the student only. Encode = single-query encoding latency. Ret/$N$ = scoring latency against $N$ candidate documents. Index/1M = index storage for 1M documents (float32 for single-vector, float16 for multi-vector). \ours shares the teacher's (Qwen3-VL-Embedding-2B) pre-computed document index. The teacher itself is excluded as it is used only for offline GPU-based document indexing.}
\label{tab:efficiency}
\begin{tabular}{llrrrlrrrr}
\toprule
Model & Params & Dim & Tok/Doc & Size & Scoring & Encode & Ret/1K & Ret/10K & Index/1M \\
\midrule
\multicolumn{10}{l}{\textit{Multi-vector VLMs (MaxSim scoring)}} \\
\quad Tomoro-8B      & 8{,}768M & 320  & 1{,}280 & 35.1\,GB & MaxSim & 8{,}225\,ms & 706\,ms   & 7{,}124\,ms & 819\,GB \\
\quad ColNomic-7B    & 7{,}829M & 128  & 1{,}000 & 31.3\,GB & MaxSim & 7{,}471\,ms & 234\,ms   & 2{,}385\,ms & 256\,GB \\
\quad ColPali        & 2{,}964M & 128  & 1{,}030 & 11.9\,GB & MaxSim & 7{,}284\,ms & 269\,ms   & 2{,}786\,ms & 264\,GB \\
\quad ColModernVBert &    259M  & 128  & 1{,}000 &  1.0\,GB & MaxSim &    183\,ms  & 242\,ms   & 2{,}553\,ms & 256\,GB \\
\midrule
\multicolumn{10}{l}{\textit{Single-vector models (Cosine scoring)}} \\
\quad DSE-Qwen2      & 2{,}209M & 1{,}536 & 1 & 8.8\,GB & Cosine & 2{,}539\,ms & 0.14\,ms & 2.3\,ms & 6.1\,GB \\
\quad SigLIP2        & 1{,}136M & 1{,}152 & 1 & 4.5\,GB & Cosine &    952\,ms  & 0.14\,ms & 2.0\,ms & 4.6\,GB \\
\midrule
\multicolumn{10}{l}{\textit{\ours (text-only student, Cosine scoring)}} \\
\quad \ours-L        &    151M  & 2{,}048 & 1 & 605\,MB & Cosine & 109\,ms & 0.20\,ms & 2.6\,ms & 8.2\,GB \\
\quad \ours-M        &    112M  & 2{,}048 & 1 & 447\,MB & Cosine & 101\,ms & 0.21\,ms & 2.5\,ms & 8.2\,GB \\
\quad \ours-S        &     69M  & 2{,}048 & 1 & 274\,MB & Cosine &  51\,ms & 0.22\,ms & 2.5\,ms & 8.2\,GB \\
\bottomrule
\end{tabular}
\end{table*}

\section{Experimental Setup}
\label{sec:setup}

We evaluate \ours on the ViDoRe benchmark against 11 baselines spanning four model categories, followed by systematic ablation of the distillation objective (\S\ref{sec:ablation}).

\subsection{Datasets and Evaluation}

We evaluate on the full public ViDoRe benchmark \cite{faysse2025colpaliefficientdocumentretrieval,macé2025vidorebenchmarkv2raising,loison-etal-2026-vidore}, comprising 22 datasets across three versions (Appendix~\ref{app:vidore-benchmark}): v1 (10 datasets: DocVQA, ArXivQA, InfoVQA, TabFQuAD, TatDQA, ShiftProject, and four SyntheticDocQA domains), v2 (4 datasets: ESG reports, biomedical lectures, economics reports, and human-labeled ESG reports), and v3 (8 datasets: finance with English and French corpora, HR, energy, industrial, pharmaceutical, physics, computer science).
We report \ndcg \cite{10.1145/582415.582418} as the primary metric, averaged per benchmark version.

For training, we aggregate 726K query-document image pairs from four public sources after quality filtering and case-insensitive deduplication:
VisRAG-Synthetic \cite{yu2025visragvisionbasedretrievalaugmentedgeneration} (239K, 32.9\%),
ColPali training set \cite{faysse2025colpaliefficientdocumentretrieval} (111K, 15.3\%),
VisRAG-InDomain \cite{yu2025visragvisionbasedretrievalaugmentedgeneration} (96K, 13.2\%),
and VDR-Multilingual \cite{cimolai2025vdr} (en/es/it/de/fr) (280K, 38.6\%).
We hold out 2\% via stratified sampling for validation (14.5K pairs), yielding 711K training pairs (Appendix~\ref{app:training-data}).
The validation set is used for model selection (best checkpoint by validation loss).

\subsection{Baselines}

We select 11 baselines that represent the full spectrum of current VDR approaches, from large multi-vector VLMs to lightweight vision-native encoders, to contextualize \ours's efficiency--quality tradeoff:
(1)~\textbf{Multi-vector VLMs} with MaxSim scoring: Tomoro-8B/4B \cite{tomoro2025colqwen3}, ColNomic-7B \cite{nomic2025colnomic}, ColPali \cite{faysse2025colpaliefficientdocumentretrieval}, ColModernVBert \cite{teiletche2025modernvbertsmallervisualdocument};
(2)~\textbf{Single-vector VLMs}: our teacher Qwen3-VL-Embedding-2B \cite{li2026qwen3vlembeddingqwen3vlrerankerunifiedframework} and DSE-Qwen2 \cite{ma2024unifyingmultimodalretrievaldocument};
(3)~\textbf{Vision-native encoders}: SigLIP2 \cite{tschannen2025siglip2multilingualvisionlanguage}, Jina-CLIP \cite{koukounas2024jinaclipclipmodel}, BiModernVBert \cite{teiletche2025modernvbertsmallervisualdocument}. (4)~\textbf{Generate-then-encode}: SERVAL \cite{nguyen-etal-2025-serval}, whose ViDoRe v2 scores we recompute from its published per-dataset results (see Table~\ref{tab:main} footnote).
Models in categories (1)--(2) and BiModernVBert are fine-tuned for document retrieval; SigLIP2 and Jina-CLIP are general-purpose contrastive models used zero-shot. All baselines except SERVAL are evaluated by us under identical conditions; SERVAL's scores are taken from its own paper.

\subsection{Implementation Details}

The VLM teacher is Qwen3-VL-Embedding-2B \cite{li2026qwen3vlembeddingqwen3vlrerankerunifiedframework} (built on Qwen3-VL \cite{bai2025qwen3vltechnicalreport}), producing $d\!=\!2048$-dimensional embeddings.
We train three student variants of increasing capacity:
\ours-S (DistilBERT \cite{sanh2020distilbertdistilledversionbert}, 66M+2M projector = 69M),
\ours-M (BERT-base \cite{devlin-etal-2019-bert}, 110M+2M = 112M), and
\ours-L (ModernBERT-base \cite{warner-etal-2025-smarter}, 149M+2M = 151M).
Each uses a two-layer MLP projector ($768 \!\to\! 768 \!\to\! 2048$) to match the teacher's embedding space.

All experiments are conducted on NVIDIA H200 GPUs (141\,GB HBM3e); all GPU-hour figures in this paper refer to this hardware.
Training uses OneCycleLR scheduling (peak lr$=$2e-4, 3\% warmup), batch size 256 with gradient accumulation 4 (effective 1024), for 20 epochs (${\sim}$13.9K steps).
Training takes 10--12 hours per model on a single GPU (10.1h for \ours-S, 10.5h for \ours-M, 11.7h for \ours-L).
Since the alignment objective is purely pointwise and query-centric, the total training cost for our best \ours model including pre-caching teacher query embeddings (${\sim}$1 GPU-hour via text-mode inference), is under 13 GPU-hours. Ranking-based ablation variants additionally require cached document embeddings; a full cost comparison is in Appendix~\ref{app:training-cost}.

\section{Main Results}
\label{sec:results}

\subsection{Retrieval Performance against Heavyweight VLMs}

Table~\ref{tab:main} presents \ndcg across the three ViDoRe benchmark versions (per-dataset breakdowns in Appendix~\ref{app:per-dataset}).

\paragraph{\ours-S approaches VLM-level quality.}
Our smallest variant, \ours-S (DistilBERT, 69M), achieves 82.2/60.5/43.5 on v1/v2/v3, retaining 92.4\% of teacher performance with 29$\times$ fewer parameters.
With multilingual query augmentation (\ours-S-Multi; \S\ref{sec:multilingual-augment}), retention rises to 95.1\% (82.2/61.9/46.5), making the 69M text-only student competitive with the 2B VLM teacher.
The larger backbones \ours-M (112M) and \ours-L (151M) offer only marginal improvements over \ours-S, confirming that the query embedding task does not require extensive model capacity.

\paragraph{Text-only students outperform VLM baselines on harder benchmarks.}
On the more challenging v2 and v3, all \ours variants surpass both ColPali (3B, multi-vector) and DSE-Qwen2 (2B, single-vector), despite being text-only models with 32$\times$ fewer parameters than DSE-Qwen2.
\ours-S-Multi achieves the highest \ndcg among all student variants on v3 (46.5, +4.5 over ColPali); \ours-M leads on v2 (62.2, +6.5 over DSE-Qwen2).

\subsection{Extreme Efficiency and Deployment Cost}

Table~\ref{tab:efficiency} summarizes deployment costs.
The efficiency gains of \ours come from two sources: the distilled student itself (query encoding latency and model size), and the inherited single-vector representation from the teacher (retrieval scoring and index storage).
All latency measurements are collected on a single CPU thread (AMD EPYC 9474F, batch size 1).

\paragraph{Query encoding latency and model size (distillation advantage).}
\ours-S encodes a query in 51\,ms on CPU---50$\times$ faster than DSE-Qwen2 (2.5\,s) and 143$\times$ faster than ColPali (7.3\,s); even \ours-L completes in 109\,ms.
The \ours-S checkpoint is 274\,MB, compared to 11.9\,GB for ColPali and 35.1\,GB for Tomoro-8B, enabling deployment on edge devices without GPU memory.

\paragraph{Inherited efficiency from single-vector teacher.}
The remaining advantages---single-vector cosine scoring (2.5\,ms per 10K documents vs.\ 7.1\,s for MaxSim) and compact index storage (8.2\,GB vs.\ 264--819\,GB per 1M pages)---are inherited from the teacher's single-vector architecture; our contribution is making the query encoder small enough to exploit them in practice.

\section{Ablation and Analysis}
\label{sec:ablation}

Our ablation spans 6 loss configurations $\times$ 3 student backbones $\times$ 3 benchmarks = 54 evaluation points, all trained to convergence under the identical settings described in \S\ref{sec:setup} (711K pairs, 20 epochs).
The only controlled variable is the loss function.
Note that ranking-based and InfoNCE variants require pre-cached teacher document embeddings for computing in-batch similarity distributions ($B\!=\!256$ in-batch negatives); the pure alignment objective does not.

\subsection{The Monotonic Superiority of Spatial Alignment}

We compare six distillation objectives spanning the full spectrum from pure ranking ($\lambda_a\!=\!0, \lambda_r\!=\!1$) to pure alignment ($\lambda_a\!=\!1, \lambda_r\!=\!0$), plus a hard-label InfoNCE baseline (Table~\ref{tab:loss}).
The combined loss is $\mathcal{L} = \lambda_a \mathcal{L}_\text{align} + \lambda_r \mathcal{L}_\text{rank}$, where $\mathcal{L}_\text{rank}$ is KL-divergence over in-batch similarity distributions:
\begin{align}
    \mathbf{p}_t &= \mathrm{softmax}\!\big(\mathbf{v}^Q_t {\mathbf{V}^D}^\top / \tau_t\big), \label{eq:rank-pt} \\
    \mathbf{p}_s &= \mathrm{softmax}\!\big(\mathbf{v}^Q_s {\mathbf{V}^D}^\top / \tau_s\big), \label{eq:rank-ps} \\
    \mathcal{L}_\text{rank} &= D_\text{KL}(\mathbf{p}_t \| \mathbf{p}_s) \label{eq:rank}
\end{align}
where $\mathbf{V}^D \in \mathbb{R}^{B \times d}$ is the matrix of in-batch teacher document embeddings, and $\tau_t\!=\!0.07$, $\tau_s\!=\!0.05$ are temperature parameters (the softer teacher distribution encourages the student to preserve relative ranking structure).
The InfoNCE baseline replaces soft teacher distributions with hard one-hot labels, where $\mathbf{v}^D_+$ is the positive document embedding:
\begin{equation}
    \mathcal{L}_\text{InfoNCE} = -\log\frac{\exp(\mathbf{v}^Q_s \cdot \mathbf{v}^D_+ / \tau_s)}{\sum_j \exp(\mathbf{v}^Q_s \cdot \mathbf{v}^D_j / \tau_s)}
    \label{eq:infonce}
\end{equation}

The result is consistent: as the alignment weight increases relative to ranking, \ndcg improves monotonically in the 3-backbone average on all three benchmarks, with consistent trends per backbone (Appendix~\ref{app:loss-backbone}).
Pure alignment outperforms pure ranking by +1.1/+4.0/+2.5 on v1/v2/v3.
This is notable because ranking-based losses (KL-divergence, MarginMSE) are the prevailing choice in retrieval distillation \cite{hofstätter2021efficientlyteachingeffectivedense,hofstätter2021improvingefficientneuralranking}.
We conjecture that alignment's advantage stems from the high quality of our teacher's embedding space: when the teacher provides well-structured coordinates, direct spatial alignment exploits richer geometric signal than relative ranking alone.
Supporting this, we find that teacher quality is the strongest predictor of distillation success ($r\!=\!+0.607$), while student--teacher cosine similarity shows near-zero correlation with retention ($r\!=\!+0.094$; Appendix~\ref{app:residual-factors})---suggesting that the geometric structure of the space matters more than pointwise proximity.

\begin{table}[t]
\centering
\small
\setlength{\tabcolsep}{8pt}
\caption{Loss ablation (\ndcg $\times$100, averaged over 3 backbones). $\lambda_a$, $\lambda_r$ = weights for $\mathcal{L}_\text{align}$ and $\mathcal{L}_\text{rank}$. All trained under identical settings (20 epochs, lr=2e-4).}
\label{tab:loss}
\begin{tabular}{ccrrr}
\toprule
$\lambda_a$ & $\lambda_r$ & v1 (10) & v2 (4) & v3 (8) \\
\midrule
1   & 0   & \textbf{82.2} & \textbf{61.4} & \textbf{44.1} \\
1   & 0.5 & 81.6 & 59.8 & 42.8 \\
1   & 1   & 81.5 & 59.1 & 42.5 \\
0.5 & 1   & 81.5 & 58.6 & 42.1 \\
0   & 1   & 81.1 & 57.4 & 41.6 \\
\midrule
\multicolumn{2}{c}{InfoNCE} & 71.5 & 39.8 & 30.0 \\
\bottomrule
\end{tabular}
\end{table}

\subsection{The Necessity of Soft-Label Distillation}

The InfoNCE \cite{oord2019representationlearningcontrastivepredictive} baseline uses hard one-hot labels---the student learns that query $i$ matches document $i$ and nothing else---rather than the teacher's soft ranking distribution.
The degradation is severe: $-$10.7 on v1, $-$21.6 on v2, and $-$14.1 on v3 compared to alignment (Table~\ref{tab:loss}, bottom row).

This 10--22 point gap shows that faithfully reproducing the teacher's continuous embedding geometry is substantially more informative than fitting binary relevance boundaries.
The teacher's ``dark knowledge'' \cite{hinton2015distillingknowledgeneuralnetwork} (the geometric relationships encoded in its embedding space) is the critical ingredient for cross-modal transfer.

\subsection{Data Efficiency}
\label{sec:data-efficiency}

Training on random subsets of the 711K pairs reveals strong diminishing returns (Appendix~\ref{app:data-efficiency}): at 25\% of training data (178K pairs), \ours-S already achieves 93\%/82\%/70\% retention on v1/v2/v3.
Even at 10\% (71K pairs), the model reaches 79\% retention on v1 (66.7 NDCG@5).
The v3 benchmark saturates slower than v1, likely because it demands broader cross-lingual coverage (\S\ref{sec:cross-lingual}).

\begin{table}[t]
\centering
\small
\setlength{\tabcolsep}{3.5pt}
\caption{Distillation retention by query language (\ours-S). Train\% = proportion of 711K training pairs in each language. Teacher and Student are mean \ndcg ($\times$100). Ret.\ = Student\,/\,Teacher. All 22 ViDoRe datasets pooled (19{,}537 unique queries after exact-duplicate removal).}
\label{tab:lang-retention}
\begin{tabular}{lrrrrr}
\toprule
Language & Train\% & \#Queries & Teacher & Student & Ret. \\
\midrule
English    & 68.7 & 6{,}237 & 64.0 & 60.3 & 94.3\% \\
French     &  7.6 & 3{,}074 & 56.2 & 51.7 & 92.1\% \\
Italian    &  7.6 & 2{,}419 & 49.0 & 44.1 & 90.0\% \\
Spanish    &  8.1 & 2{,}694 & 51.4 & 46.1 & 89.7\% \\
German     &  8.0 & 2{,}694 & 49.3 & 42.3 & 85.7\% \\
Portuguese &  0.0 & 2{,}419 & 48.7 & 36.8 & 75.6\% \\
\bottomrule
\end{tabular}
\end{table}

\subsection{Cross-Lingual Transfer: The Primary Bottleneck}
\label{sec:cross-lingual}

A natural concern with text-only distillation is whether the modality gap (text-only student vs.\ vision-language teacher) limits performance on visual documents.
We conduct a per-query analysis on \ours-S across all 22 ViDoRe datasets (19{,}537 queries), grouping every query by language to disentangle language effects from document content.

\paragraph{Language determines retention.}
Table~\ref{tab:lang-retention} aggregates retention (student \ndcg / teacher \ndcg $\times$ 100) by the six evaluation languages.
The hierarchy broadly tracks training data coverage: English (68.7\% of training data) achieves 94.3\% retention, French/Italian/Spanish (7--8\% each) 90--92\%, German (8.0\%) 85.7\%, and Portuguese (entirely absent from training) only 75.6\%.
The Pearson correlation between training data proportion and retention is $r\!=\!+0.563$ ($n\!=\!6$; suggestive but not statistically significant at this sample size).

\paragraph{Within-dataset evidence isolates the language effect.}
On the eight ViDoRe v3 multilingual subsets---where the same document corpus is queried in all six languages---English queries average 92.8\% retention versus 75.4\% for Portuguese, a 17.4~pp gap on identical corpora.
Per-dataset breakdowns (Appendix~\ref{app:lang-matrix}) show the gap is largest for dense technical terminology and smallest for notation-heavy scientific content.

\paragraph{The modality gap is not the bottleneck.}
For English queries, \ours-S retains 94.3\% of teacher quality across 20 datasets spanning diverse visual document types---infographics, financial tables, scientific diagrams, and dense text---demonstrating that a text-only student can effectively retrieve visual documents through spatial alignment alone.
The dominant bottleneck is cross-lingual transfer: \ours-S uses DistilBERT, a primarily English-trained backbone, which limits its ability to align non-English query embeddings with the teacher's multilingual space. Since alignment training is query-centric and requires no document images, this gap can be addressed cheaply via query-only data augmentation (\S\ref{sec:multilingual-augment}).

Beyond language, we find that teacher quality ($r\!=\!+0.607$) and corpus size ($r\!=\!-0.566$) predict the remaining retention variation on English-only queries, while direct student--teacher cosine similarity shows near-zero correlation ($r\!=\!+0.094$); full analysis is in Appendix~\ref{app:residual-factors}.

\subsection{Validating Multilingual Augmentation}
\label{sec:multilingual-val}

Table~\ref{tab:multilingual-aug} validates the multilingual augmentation described in \S\ref{sec:multilingual-augment} by breaking down \ours-S-Multi's improvement by query language.
The gains concentrate precisely where the cross-lingual analysis above predicted: Portuguese (previously absent from training) gains +9.3 NDCG, while English shows zero regression.
After augmentation, all six languages achieve $\ge$92\% retention---the maximum cross-lingual gap narrows from 18.6~pp to 2.7~pp.
On the benchmark level (Table~\ref{tab:main}, \ours-S-Multi), ViDoRe v1 performance is preserved (82.2 $\to$ 82.2); the multilingual v2 and v3 gain +1.4 and +3.0 respectively.
These results confirm that the primary bottleneck for \ours is cross-lingual transfer, not cross-modal transfer, and that it can be resolved cheaply through query-only data augmentation.

\begin{table}[t]
\centering
\small
\setlength{\tabcolsep}{2pt}
\caption{Effect of multilingual query augmentation by language (\ndcg $\times$100). Original = \ours-S trained on 711K pairs; +Aug = retrained on 1.49M pairs with translated queries. $\Delta$ = +Aug $-$ Original. Ret.\ = +Aug\,/\,Teacher. Same 19{,}537 queries as Table~\ref{tab:lang-retention}.}
\label{tab:multilingual-aug}
\begin{tabular}{lrrrrrr}
\toprule
Language & \#Queries & Teacher & Original & +Aug & $\Delta$ & Ret. \\
\midrule
English    & 6{,}237 & 64.0 & 60.3 & 60.3 & +0.0 & 94.3\% \\
French     & 3{,}074 & 56.2 & 51.7 & 53.2 & +1.5 & 94.7\% \\
Italian    & 2{,}419 & 49.0 & 44.1 & 45.7 & +1.6 & 93.3\% \\
Spanish    & 2{,}694 & 51.4 & 46.1 & 47.8 & +1.8 & 93.1\% \\
German     & 2{,}694 & 49.3 & 42.3 & 45.4 & +3.1 & 92.0\% \\
Portuguese & 2{,}419 & 48.7 & 36.8 & 46.1 & +9.3 & 94.6\% \\
\bottomrule
\end{tabular}
\end{table}

\section{Conclusion}
\label{sec:conclusion}

We presented \ours, a framework for asymmetric cross-modal distillation that enables efficient visual document retrieval.
Given a high-quality VLM teacher, the simplest possible objective---pointwise cosine alignment---consistently outperforms ranking-based and contrastive alternatives, while eliminating document representations and negative sampling from training.
The query-centric nature of alignment makes the primary bottleneck cross-lingual rather than cross-modal, which we resolve cheaply via translated query augmentation, narrowing a large language gap without additional image processing.
With this augmentation, \ours-S-Multi (DistilBERT, 69M) retains 95.1\% of teacher quality and outperforms DSE-Qwen2 (2B) on v2 and v3 with 32$\times$ fewer parameters, 50$\times$ lower query latency, and 32--100$\times$ less index storage---all at a training cost under 13 GPU-hours.

\section*{Limitations}

\ours specialises the query path for text.
The teacher, Qwen3-VL-Embedding-2B, accepts image queries and interleaved image--text queries; the distilled student does not, and cannot be made to without reintroducing a vision encoder.
This is a genuine capability loss, not only a scope choice.
It is invisible on ViDoRe, whose 22 datasets all pair a short text query with a page-image corpus, and it matches how visual document retrieval is deployed in enterprise search and RAG today, but any application that needs an image as the query must keep the full VLM on the query side.
Distilling a multi-modal query path into a compact student is left to future work.

\ours inherits not only the teacher's embedding quality but also its representation form: one vector per page, scored by dot product.
This is the structural reason it does not match late-interaction systems on the hardest benchmark, where Tomoro-8B reaches 59.0 on ViDoRe v3 against 46.5 for \ours-S-Multi.
Student-side capacity does not close that gap, because the information a multi-vector index keeps per page was already discarded when the teacher pooled it.
The trade is deliberate: the same pooling is what buys the 100$\times$ smaller index and the 2.5\,ms scoring latency in Table~\ref{tab:efficiency}.
Distilling a multi-vector teacher into a token-level student, trading that index back for quality, is a concrete direction we have not explored.

This work does not explore reducing the offline indexing cost, which still requires the full 2B VLM teacher to encode every document image; teacher compression and progressive indexing remain future work.
We evaluate exclusively on first-stage ranking; whether the framework generalises to re-ranking or to text-only corpora is untested.
Our multilingual query augmentation relies on lightweight machine translation models, which may introduce semantic shifts or struggle with domain-specific terminology (for example in finance or physics); higher-fidelity LLM translation is a natural extension. 

The 512-token limit of the DistilBERT backbone is not binding on current benchmarks: across all 19{,}609 ViDoRe queries the longest is 182 tokens and none is truncated (Appendix~\ref{app:query-length}). Retrieval settings with paragraph-length or multi-hop queries fall outside what these benchmarks measure, and whether a 512-token student suffices there is untested.

\section*{Acknowledgments}

This work was carried out in AiWo: Human-Centric, AI-Enabled Collaborative Fieldwork
Operations, a Business Finland co-innovation programme under grant agreement No.~69/31/2025.
The computations were performed using computer resources provided by the Aalto University
School of Science ``Science-IT'' project.

We used Claude Code as a coding and writing assistant: implementing training
and evaluation scripts, preparing and analysing data, and assisting with the language of
the manuscript. The research questions, the experimental design, the interpretation of
results, and all claims in this paper are the authors' own; every reported number was
produced by running the released code and verified by the authors.

\bibliography{custom}

\begin{thebibliography}{39}
\providecommand{\natexlab}[1]{#1}

\bibitem[{Bai et~al.(2025)Bai, Cai, Chen, Chen, Chen, Cheng, Deng, Ding, Gao, Ge, Ge, Guo, Huang, Huang, Huang, Hui, Jiang, Li, Li, Li, Li, Lin, Lin, Liu, Liu, Liu, Liu, Liu, Liu, Lu, Luo, Lv, Men, Meng, Ren, Ren, Song, Sun, Tang, Tu, Wan, Wang, Wang, Wang, Wang, Xie, Xu, Xu, Xu, Yang, Yang, Yang, Yang, Yu, Zhang, Zhang, Zhang, Zheng, Zhong, Zhou, Zhou, Zhou, Zhu, and Zhu}]{bai2025qwen3vltechnicalreport}
Shuai Bai, Yuxuan Cai, Ruizhe Chen, Keqin Chen, Xionghui Chen, Zesen Cheng, Lianghao Deng, Wei Ding, Chang Gao, Chunjiang Ge, Wenbin Ge, Zhifang Guo, Qidong Huang, Jie Huang, Fei Huang, Binyuan Hui, Shutong Jiang, Zhaohai Li, Mingsheng Li, and 45 others. 2025.
\newblock \href {https://arxiv.org/abs/2511.21631} {Qwen3-vl technical report}.
\newblock \emph{Preprint}, arXiv:2511.21631.

\bibitem[{Cimolai and Markewich(2025)}]{cimolai2025vdr}
Marco Cimolai and Logan Markewich. 2025.
\newblock Visual document retrieval goes multilingual.
\newblock Hugging Face Blog.
\newblock \url{https://huggingface.co/blog/vdr-2b-multilingual}.

\bibitem[{Devlin et~al.(2019)Devlin, Chang, Lee, and Toutanova}]{devlin-etal-2019-bert}
Jacob Devlin, Ming-Wei Chang, Kenton Lee, and Kristina Toutanova. 2019.
\newblock \href {https://doi.org/10.18653/v1/N19-1423} {{BERT}: Pre-training of deep bidirectional transformers for language understanding}.
\newblock In \emph{Proceedings of the 2019 Conference of the North {A}merican Chapter of the Association for Computational Linguistics: Human Language Technologies, Volume 1 (Long and Short Papers)}, pages 4171--4186, Minneapolis, Minnesota. Association for Computational Linguistics.

\bibitem[{Faysse et~al.(2025)Faysse, Sibille, Wu, Omrani, Viaud, Hudelot, and Colombo}]{faysse2025colpaliefficientdocumentretrieval}
Manuel Faysse, Hugues Sibille, Tony Wu, Bilel Omrani, Gautier Viaud, Céline Hudelot, and Pierre Colombo. 2025.
\newblock \href {https://arxiv.org/abs/2407.01449} {Colpali: Efficient document retrieval with vision language models}.
\newblock \emph{Preprint}, arXiv:2407.01449.

\bibitem[{Hinton et~al.(2015)Hinton, Vinyals, and Dean}]{hinton2015distillingknowledgeneuralnetwork}
Geoffrey Hinton, Oriol Vinyals, and Jeff Dean. 2015.
\newblock \href {https://arxiv.org/abs/1503.02531} {Distilling the knowledge in a neural network}.
\newblock \emph{Preprint}, arXiv:1503.02531.

\bibitem[{Hofstätter et~al.(2021{\natexlab{a}})Hofstätter, Althammer, Schröder, Sertkan, and Hanbury}]{hofstätter2021improvingefficientneuralranking}
Sebastian Hofstätter, Sophia Althammer, Michael Schröder, Mete Sertkan, and Allan Hanbury. 2021{\natexlab{a}}.
\newblock \href {https://arxiv.org/abs/2010.02666} {Improving efficient neural ranking models with cross-architecture knowledge distillation}.
\newblock \emph{Preprint}, arXiv:2010.02666.

\bibitem[{Hofstätter et~al.(2021{\natexlab{b}})Hofstätter, Lin, Yang, Lin, and Hanbury}]{hofstätter2021efficientlyteachingeffectivedense}
Sebastian Hofstätter, Sheng-Chieh Lin, Jheng-Hong Yang, Jimmy Lin, and Allan Hanbury. 2021{\natexlab{b}}.
\newblock \href {https://arxiv.org/abs/2104.06967} {Efficiently teaching an effective dense retriever with balanced topic aware sampling}.
\newblock \emph{Preprint}, arXiv:2104.06967.

\bibitem[{J{\"a}rvelin and Kek{\"a}l{\"a}inen(2002)}]{10.1145/582415.582418}
Kalervo J{\"a}rvelin and Jaana Kek{\"a}l{\"a}inen. 2002.
\newblock \href {https://doi.org/10.1145/582415.582418} {Cumulated gain-based evaluation of ir techniques}.
\newblock \emph{ACM Trans. Inf. Syst.}, 20(4):422–446.

\bibitem[{Karpukhin et~al.(2020)Karpukhin, Oğuz, Min, Lewis, Wu, Edunov, Chen, and tau Yih}]{karpukhin2020densepassageretrievalopendomain}
Vladimir Karpukhin, Barlas Oğuz, Sewon Min, Patrick Lewis, Ledell Wu, Sergey Edunov, Danqi Chen, and Wen tau Yih. 2020.
\newblock \href {https://arxiv.org/abs/2004.04906} {Dense passage retrieval for open-domain question answering}.
\newblock \emph{Preprint}, arXiv:2004.04906.

\bibitem[{Khattab and Zaharia(2020)}]{khattab2020colbertefficienteffectivepassage}
Omar Khattab and Matei Zaharia. 2020.
\newblock \href {https://arxiv.org/abs/2004.12832} {Colbert: Efficient and effective passage search via contextualized late interaction over bert}.
\newblock \emph{Preprint}, arXiv:2004.12832.

\bibitem[{Kim et~al.(2023)Kim, Rawat, Zaheer, Jayasumana, Sadhanala, Jitkrittum, Menon, Fergus, and Kumar}]{kim2023embeddistillgeometricknowledgedistillation}
Seungyeon Kim, Ankit~Singh Rawat, Manzil Zaheer, Sadeep Jayasumana, Veeranjaneyulu Sadhanala, Wittawat Jitkrittum, Aditya~Krishna Menon, Rob Fergus, and Sanjiv Kumar. 2023.
\newblock \href {https://arxiv.org/abs/2301.12005} {Embeddistill: A geometric knowledge distillation for information retrieval}.
\newblock \emph{Preprint}, arXiv:2301.12005.

\bibitem[{Koukounas et~al.(2024)Koukounas, Mastrapas, Günther, Wang, Martens, Mohr, Sturua, Akram, Martínez, Ognawala, Guzman, Werk, Wang, and Xiao}]{koukounas2024jinaclipclipmodel}
Andreas Koukounas, Georgios Mastrapas, Michael Günther, Bo~Wang, Scott Martens, Isabelle Mohr, Saba Sturua, Mohammad~Kalim Akram, Joan~Fontanals Martínez, Saahil Ognawala, Susana Guzman, Maximilian Werk, Nan Wang, and Han Xiao. 2024.
\newblock \href {https://arxiv.org/abs/2405.20204} {Jina clip: Your clip model is also your text retriever}.
\newblock \emph{Preprint}, arXiv:2405.20204.

\bibitem[{Li et~al.(2026)Li, Zhang, Long, Chen, Song, Bai, Yang, Xie, Yang, Liu, Zhou, and Lin}]{li2026qwen3vlembeddingqwen3vlrerankerunifiedframework}
Mingxin Li, Yanzhao Zhang, Dingkun Long, Keqin Chen, Sibo Song, Shuai Bai, Zhibo Yang, Pengjun Xie, An~Yang, Dayiheng Liu, Jingren Zhou, and Junyang Lin. 2026.
\newblock \href {https://arxiv.org/abs/2601.04720} {Qwen3-vl-embedding and qwen3-vl-reranker: A unified framework for state-of-the-art multimodal retrieval and ranking}.
\newblock \emph{Preprint}, arXiv:2601.04720.

\bibitem[{Liu et~al.(2026)Liu, Hu, Zhang, and Xiao}]{liu2026structuralanchorpruningtrainingfree}
Zhuchenyang Liu, Ziyu Hu, Yao Zhang, and Yu~Xiao. 2026.
\newblock \href {https://arxiv.org/abs/2601.20107} {Structural anchor pruning: Training-free multi-vector compression for visual document retrieval}.
\newblock \emph{Preprint}, arXiv:2601.20107.

\bibitem[{Loison et~al.(2026)Loison, Mac{\'e}, Edy, Xing, Balough, Moreira, Liu, Faysse, Hudelot, and Viaud}]{loison-etal-2026-vidore}
Ant{\'o}nio Loison, Quentin Mac{\'e}, Antoine Edy, Victor Xing, Tom Balough, Gabriel de Souza~P. Moreira, Bo~Liu, Manuel Faysse, Celine Hudelot, and Gautier Viaud. 2026.
\newblock \href {https://doi.org/10.18653/v1/2026.acl-long.755} {{V}i{D}o{R}e v3: A comprehensive evaluation of retrieval augmented generation in complex real-world scenarios}.
\newblock In \emph{Proceedings of the 64th Annual Meeting of the {A}ssociation for {C}omputational {L}inguistics (Volume 1: Long Papers)}, pages 16570--16600, San Diego, California, United States. Association for Computational Linguistics.

\bibitem[{Ltd(2025)}]{tomoro2025colqwen3}
Tomoro~AI Ltd. 2025.
\newblock \href {https://huggingface.co/TomoroAI/tomoro-colqwen3-embed-8b} {Tomoroai/tomoro-colqwen3-embed-8b}.

\bibitem[{Ma et~al.(2024)Ma, Lin, Li, Chen, and Lin}]{ma2024unifyingmultimodalretrievaldocument}
Xueguang Ma, Sheng-Chieh Lin, Minghan Li, Wenhu Chen, and Jimmy Lin. 2024.
\newblock \href {https://arxiv.org/abs/2406.11251} {Unifying multimodal retrieval via document screenshot embedding}.
\newblock \emph{Preprint}, arXiv:2406.11251.

\bibitem[{Ma et~al.(2025)Ma, Li, Zang, Wu, Dong, Zhang, Cao, Duan, Wang, Cao, and Sun}]{ma-etal-2025-towards-storage}
Yubo Ma, Jinsong Li, Yuhang Zang, Xiaobao Wu, Xiaoyi Dong, Pan Zhang, Yuhang Cao, Haodong Duan, Jiaqi Wang, Yixin Cao, and Aixin Sun. 2025.
\newblock \href {https://doi.org/10.18653/v1/2025.findings-acl.1003} {Towards storage-efficient visual document retrieval: An empirical study on reducing patch-level embeddings}.
\newblock In \emph{Findings of the Association for Computational Linguistics: ACL 2025}, pages 19568--19580, Vienna, Austria. Association for Computational Linguistics.

\bibitem[{Macé et~al.(2025)Macé, Loison, and Faysse}]{macé2025vidorebenchmarkv2raising}
Quentin Macé, António Loison, and Manuel Faysse. 2025.
\newblock \href {https://arxiv.org/abs/2505.17166} {Vidore benchmark v2: Raising the bar for visual retrieval}.
\newblock \emph{Preprint}, arXiv:2505.17166.

\bibitem[{Nguyen et~al.(2025)Nguyen, Lei, Ju, and Yates}]{nguyen-etal-2025-serval}
Thong Nguyen, Yibin Lei, Jia-Huei Ju, and Andrew Yates. 2025.
\newblock \href {https://doi.org/10.18653/v1/2025.emnlp-main.1568} {{SERVAL}: Surprisingly effective zero-shot visual document retrieval powered by large vision and language models}.
\newblock In \emph{Proceedings of the 2025 Conference on Empirical Methods in Natural Language Processing}, pages 30807--30822, Suzhou, China. Association for Computational Linguistics.

\bibitem[{{Nomic AI}(2025)}]{nomic2025colnomic}
{Nomic AI}. 2025.
\newblock nomic-ai/colnomic-embed-multimodal-7b.

\bibitem[{Reddi et~al.(2021)Reddi, Kumar~Pasumarthi, Menon, Singh~Rawat, Yu, Kim, Veit, and Kumar}]{pmlr-v130-reddi21a}
Sashank Reddi, Rama Kumar~Pasumarthi, Aditya Menon, Ankit Singh~Rawat, Felix Yu, Seungyeon Kim, Andreas Veit, and Sanjiv Kumar. 2021.
\newblock \href {https://proceedings.mlr.press/v130/reddi21a.html} {Rankdistil: Knowledge distillation for ranking}.
\newblock In \emph{Proceedings of The 24th International Conference on Artificial Intelligence and Statistics}, volume 130 of \emph{Proceedings of Machine Learning Research}, pages 2368--2376. PMLR.

\bibitem[{Reimers and Gurevych(2019)}]{reimers2019sentencebertsentenceembeddingsusing}
Nils Reimers and Iryna Gurevych. 2019.
\newblock \href {https://arxiv.org/abs/1908.10084} {Sentence-bert: Sentence embeddings using siamese bert-networks}.
\newblock \emph{Preprint}, arXiv:1908.10084.

\bibitem[{Sanh et~al.(2020)Sanh, Debut, Chaumond, and Wolf}]{sanh2020distilbertdistilledversionbert}
Victor Sanh, Lysandre Debut, Julien Chaumond, and Thomas Wolf. 2020.
\newblock \href {https://arxiv.org/abs/1910.01108} {Distilbert, a distilled version of bert: smaller, faster, cheaper and lighter}.
\newblock \emph{Preprint}, arXiv:1910.01108.

\bibitem[{Sun et~al.(2025)Sun, Hou, Guo, Wang, Yang, Ni, and Zhang}]{sun-etal-2025-unveil}
Hao Sun, Yingyan Hou, Jiayan Guo, Bo~Wang, Chunyu Yang, Jinsong Ni, and Yan Zhang. 2025.
\newblock \href {https://doi.org/10.18653/v1/2025.acl-long.1166} {Unveil: Unified visual-textual integration and distillation for multi-modal document retrieval}.
\newblock In \emph{Proceedings of the 63rd Annual Meeting of the Association for Computational Linguistics (Volume 1: Long Papers)}, pages 23935--23945, Vienna, Austria. Association for Computational Linguistics.

\bibitem[{Teiletche et~al.(2025)Teiletche, Macé, Conti, Loison, Viaud, Colombo, and Faysse}]{teiletche2025modernvbertsmallervisualdocument}
Paul Teiletche, Quentin Macé, Max Conti, Antonio Loison, Gautier Viaud, Pierre Colombo, and Manuel Faysse. 2025.
\newblock \href {https://arxiv.org/abs/2510.01149} {Modernvbert: Towards smaller visual document retrievers}.
\newblock \emph{Preprint}, arXiv:2510.01149.

\bibitem[{Tiedemann and Thottingal(2020)}]{tiedemann-thottingal-2020-opus}
J{\"o}rg Tiedemann and Santhosh Thottingal. 2020.
\newblock \href {https://aclanthology.org/2020.eamt-1.61/} {{OPUS}-{MT} {--} building open translation services for the world}.
\newblock In \emph{Proceedings of the 22nd Annual Conference of the European Association for Machine Translation}, pages 479--480, Lisboa, Portugal. European Association for Machine Translation.

\bibitem[{Tschannen et~al.(2025)Tschannen, Gritsenko, Wang, Naeem, Alabdulmohsin, Parthasarathy, Evans, Beyer, Xia, Mustafa, Hénaff, Harmsen, Steiner, and Zhai}]{tschannen2025siglip2multilingualvisionlanguage}
Michael Tschannen, Alexey Gritsenko, Xiao Wang, Muhammad~Ferjad Naeem, Ibrahim Alabdulmohsin, Nikhil Parthasarathy, Talfan Evans, Lucas Beyer, Ye~Xia, Basil Mustafa, Olivier Hénaff, Jeremiah Harmsen, Andreas Steiner, and Xiaohua Zhai. 2025.
\newblock \href {https://arxiv.org/abs/2502.14786} {Siglip 2: Multilingual vision-language encoders with improved semantic understanding, localization, and dense features}.
\newblock \emph{Preprint}, arXiv:2502.14786.

\bibitem[{van~den Oord et~al.(2019)van~den Oord, Li, and Vinyals}]{oord2019representationlearningcontrastivepredictive}
Aaron van~den Oord, Yazhe Li, and Oriol Vinyals. 2019.
\newblock \href {https://arxiv.org/abs/1807.03748} {Representation learning with contrastive predictive coding}.
\newblock \emph{Preprint}, arXiv:1807.03748.

\bibitem[{Vujanic and R{\"u}ckstie{\ss}(2026)}]{vujanic-ruckstiess-2026-leaf}
Robin Vujanic and Thomas R{\"u}ckstie{\ss}. 2026.
\newblock \href {https://doi.org/10.18653/v1/2026.acl-long.2008} {{LEAF}: Knowledge distillation of text embedding models with teacher-aligned representations}.
\newblock In \emph{Proceedings of the 64th Annual Meeting of the {A}ssociation for {C}omputational {L}inguistics (Volume 1: Long Papers)}, pages 43362--43383, San Diego, California, United States. Association for Computational Linguistics.

\bibitem[{Wang et~al.(2024)Wang, Bai, Tan, Wang, Fan, Bai, Chen, Liu, Wang, Ge, Fan, Dang, Du, Ren, Men, Liu, Zhou, Zhou, and Lin}]{wang2024qwen2vlenhancingvisionlanguagemodels}
Peng Wang, Shuai Bai, Sinan Tan, Shijie Wang, Zhihao Fan, Jinze Bai, Keqin Chen, Xuejing Liu, Jialin Wang, Wenbin Ge, Yang Fan, Kai Dang, Mengfei Du, Xuancheng Ren, Rui Men, Dayiheng Liu, Chang Zhou, Jingren Zhou, and Junyang Lin. 2024.
\newblock \href {https://arxiv.org/abs/2409.12191} {Qwen2-vl: Enhancing vision-language model's perception of the world at any resolution}.
\newblock \emph{Preprint}, arXiv:2409.12191.

\bibitem[{Wang and Hong(2023)}]{wang-hong-2023-query}
Yuxuan Wang and Lyu Hong. 2023.
\newblock \href {https://doi.org/10.18653/v1/2023.sustainlp-1.23} {Query encoder distillation via embedding alignment is a strong baseline method to boost dense retriever online efficiency}.
\newblock In \emph{Proceedings of the Fourth Workshop on Simple and Efficient Natural Language Processing (SustaiNLP)}, pages 290--298, Toronto, Canada (Hybrid). Association for Computational Linguistics.

\bibitem[{Warner et~al.(2025)Warner, Chaffin, Clavi{\'e}, Weller, Hallstr{\"o}m, Taghadouini, Gallagher, Biswas, Ladhak, Aarsen, Adams, Howard, and Poli}]{warner-etal-2025-smarter}
Benjamin Warner, Antoine Chaffin, Benjamin Clavi{\'e}, Orion Weller, Oskar Hallstr{\"o}m, Said Taghadouini, Alexis Gallagher, Raja Biswas, Faisal Ladhak, Tom Aarsen, Griffin~Thomas Adams, Jeremy Howard, and Iacopo Poli. 2025.
\newblock \href {https://doi.org/10.18653/v1/2025.acl-long.127} {Smarter, better, faster, longer: A modern bidirectional encoder for fast, memory efficient, and long context finetuning and inference}.
\newblock In \emph{Proceedings of the 63rd Annual Meeting of the Association for Computational Linguistics (Volume 1: Long Papers)}, pages 2526--2547, Vienna, Austria. Association for Computational Linguistics.

\bibitem[{Wu et~al.(2023)Wu, Peng, Zhou, Xiao, Liu, Yuan, Xuan, Valenzuela, Xi, Chen, Wang, Chao, and Hu}]{wu2023tinyclipclipdistillationaffinity}
Kan Wu, Houwen Peng, Zhenghong Zhou, Bin Xiao, Mengchen Liu, Lu~Yuan, Hong Xuan, Michael Valenzuela, Xi, Chen, Xinggang Wang, Hongyang Chao, and Han Hu. 2023.
\newblock \href {https://arxiv.org/abs/2309.12314} {Tinyclip: Clip distillation via affinity mimicking and weight inheritance}.
\newblock \emph{Preprint}, arXiv:2309.12314.

\bibitem[{Xiao et~al.(2026)Xiao, Ma, Gu, cheng Jason~Chen, Chen, Ordonez, and Mohan}]{xiao2026metaembedscalingmultimodalretrieval}
Zilin Xiao, Qi~Ma, Mengting Gu, Chun cheng Jason~Chen, Xintao Chen, Vicente Ordonez, and Vijai Mohan. 2026.
\newblock \href {https://arxiv.org/abs/2509.18095} {Metaembed: Scaling multimodal retrieval at test-time with flexible late interaction}.
\newblock \emph{Preprint}, arXiv:2509.18095.

\bibitem[{Yan et~al.(2025)Yan, Xu, Zou, Liu, Kwok, and Hu}]{yan2025docprunerstorageefficientframeworkmultivector}
Yibo Yan, Guangwei Xu, Xin Zou, Shuliang Liu, James Kwok, and Xuming Hu. 2025.
\newblock \href {https://arxiv.org/abs/2509.23883} {Docpruner: A storage-efficient framework for multi-vector visual document retrieval via adaptive patch-level embedding pruning}.
\newblock \emph{Preprint}, arXiv:2509.23883.

\bibitem[{Yang et~al.(2024)Yang, An, Huang, Bi, Yu, Yang, Diao, and Xu}]{yang2024clipkdempiricalstudyclip}
Chuanguang Yang, Zhulin An, Libo Huang, Junyu Bi, Xinqiang Yu, Han Yang, Boyu Diao, and Yongjun Xu. 2024.
\newblock \href {https://arxiv.org/abs/2307.12732} {Clip-kd: An empirical study of clip model distillation}.
\newblock \emph{Preprint}, arXiv:2307.12732.

\bibitem[{Yu et~al.(2025)Yu, Tang, Xu, Cui, Ran, Yan, Liu, Wang, Han, Liu, and Sun}]{yu2025visragvisionbasedretrievalaugmentedgeneration}
Shi Yu, Chaoyue Tang, Bokai Xu, Junbo Cui, Junhao Ran, Yukun Yan, Zhenghao Liu, Shuo Wang, Xu~Han, Zhiyuan Liu, and Maosong Sun. 2025.
\newblock \href {https://arxiv.org/abs/2410.10594} {Visrag: Vision-based retrieval-augmented generation on multi-modality documents}.
\newblock \emph{Preprint}, arXiv:2410.10594.

\bibitem[{Zhou et~al.(2024)Zhou, Liu, Xiao, Zhao, and Xiong}]{zhou-etal-2024-vista}
Junjie Zhou, Zheng Liu, Shitao Xiao, Bo~Zhao, and Yongping Xiong. 2024.
\newblock \href {https://doi.org/10.18653/v1/2024.acl-long.175} {{VISTA}: Visualized text embedding for universal multi-modal retrieval}.
\newblock In \emph{Proceedings of the 62nd Annual Meeting of the Association for Computational Linguistics (Volume 1: Long Papers)}, pages 3185--3200, Bangkok, Thailand. Association for Computational Linguistics.

\end{thebibliography}

\appendix

\section{ViDoRe Benchmark Details}
\label{app:vidore-benchmark}

The Visual Document Retrieval (ViDoRe) benchmark comprises three progressively challenging versions.
Table~\ref{tab:vidore-versions} provides a cross-version summary; Table~\ref{tab:vidore-datasets} lists all 22 evaluation datasets.

\paragraph{v1 \cite{faysse2025colpaliefficientdocumentretrieval}.}
Introduced alongside ColPali, v1 contains 10 datasets in English and French.
Five are sourced from established visual QA benchmarks (DocVQA, ArXivQA, InfoVQA, TatDQA, TabFQuAD) with human-authored queries; five use queries generated by Claude-3 Sonnet over curated document collections.
With state-of-the-art models exceeding 90 NDCG@5, v1 is approaching saturation.

\paragraph{v2 \cite{macé2025vidorebenchmarkv2raising}.}
Designed to address v1's saturation, v2 introduces 4 datasets with queries in up to 4 languages (English, French, Spanish, German).
Queries are generated through a blind contextual process---annotators receive only document metadata, not full pages---reducing extractive bias and requiring cross-document reasoning.
The fourth dataset (ESG Reports Human-Labeled) provides expert human annotations over the same ESG corpus.
Best models score 59--65 NDCG@5.

\paragraph{v3 \cite{loison-etal-2026-vidore}.}
The most comprehensive version, v3 provides 8 public datasets across 6 languages (adding Italian and Portuguese).
Annotation involved approximately 12{,}000 person-hours, combining LLM-synthesized queries with extensive expert review.
Each query includes page-level relevancy rankings, bounding box annotations, and multilingual translations.
The 8 domains span enterprise scenarios from finance to physics, with document corpora in English or French.

\begin{table}[t]
\centering
\small
\setlength{\tabcolsep}{1pt}
\caption{Cross-version comparison of the ViDoRe benchmark.}
\label{tab:vidore-versions}
\begin{tabular}{lccc}
\toprule
 & v1 & v2 & v3 \\
\midrule
Datasets & 10 & 4 & 8 \\
Query languages & 2 & 4 & 6 \\
Query source & VQA + LLM & LLM + human & LLM + expert \\
SOTA \ndcg & $>$90 & 59--65 & 57--59 \\
\bottomrule
\end{tabular}
\end{table}

\begin{table*}[t]
\centering
\small
\setlength{\tabcolsep}{4pt}
\caption{All 22 ViDoRe evaluation datasets used in this work. Doc = document corpus language. Query = languages in which queries are provided. Source: H = human-authored (from VQA benchmarks or expert annotation); L = LLM-generated; L+H = LLM-generated with human review. Six languages = EN/FR/ES/DE/IT/PT.}
\label{tab:vidore-datasets}
\begin{tabular}{lclccl}
\toprule
Dataset & Ver. & Document Domain & Doc & Query & Source \\
\midrule
DocVQA & v1 & Industrial documents & EN & EN & H \\
ArXivQA & v1 & Scientific papers & EN & EN & H \\
InfoVQA & v1 & Infographics & EN & EN & H \\
TatDQA & v1 & Financial tables & EN & EN & H \\
TabFQuAD & v1 & Tables in French PDFs & FR & FR & H \\
SyntheticDocQA-AI & v1 & AI-related documents & EN & EN & L \\
SyntheticDocQA-Energy & v1 & Energy sector reports & EN & EN & L \\
SyntheticDocQA-Gov. & v1 & Government reports & EN & EN & L \\
SyntheticDocQA-Hlt. & v1 & Healthcare documents & EN & EN & L \\
ShiftProject & v1 & Environmental reports & FR & FR & L \\
\midrule
ESG Reports & v2 & ESG / sustainability & EN & EN/FR/ES/DE & L+H \\
Biomedical Lectures & v2 & Biomedical slides & EN & EN/FR/ES/DE & L+H \\
Economics Reports & v2 & Economics reports & EN & EN/FR/ES/DE & L+H \\
ESG Reports (Human) & v2 & ESG / sustainability & EN & EN & H \\
\midrule
Finance-EN & v3 & US annual reports & EN & 6 langs & L+H \\
Finance-FR & v3 & French annual reports & FR & 6 langs & L+H \\
Computer Science & v3 & CS textbooks & EN & 6 langs & L+H \\
HR & v3 & EU HR reports & EN & 6 langs & L+H \\
Energy & v3 & French energy reports & FR & 6 langs & L+H \\
Industrial & v3 & USAF technical orders & EN & 6 langs & L+H \\
Pharmaceutical & v3 & FDA reports & EN & 6 langs & L+H \\
Physics & v3 & French physics lectures & FR & 6 langs & L+H \\
\bottomrule
\end{tabular}
\end{table*}

\section{Training Data Details}
\label{app:training-data}

We aggregate training data from four publicly available visual document retrieval datasets, totalling 726K unique query--document image pairs after quality filtering and case-insensitive query deduplication (from 761K raw pairs).
A 2\% stratified split yields 711K training and 14.5K validation pairs.
Table~\ref{tab:train-data} provides the breakdown.

\begin{table}[t]
\centering
\small
\caption{Training data composition after deduplication. VDR-Multilingual is split by language.}
\label{tab:train-data}
\begin{tabular}{llrr}
\toprule
Source & Subset & Train & \% \\
\midrule
VisRAG-Synthetic & --- & 233{,}817 & 32.9 \\
ColPali & --- & 109{,}044 & 15.3 \\
VisRAG-InDomain & --- & 94{,}016 & 13.2 \\
\midrule
VDR-Multilingual & Spanish & 57{,}491 & 8.1 \\
 & German & 56{,}994 & 8.0 \\
 & French & 54{,}079 & 7.6 \\
 & Italian & 53{,}787 & 7.6 \\
 & English & 52{,}375 & 7.4 \\
\midrule
\textbf{Total} & & \textbf{711{,}603} & \textbf{100.0} \\
\bottomrule
\end{tabular}
\end{table}

\paragraph{VisRAG-Synthetic \cite{yu2025visragvisionbasedretrievalaugmentedgeneration}.}
This dataset contains 234K synthetically generated query--page image pairs.
Document pages are sourced from four domains:
OpenStax college-level textbooks (10K pages),
ICML 2023 papers (5K pages),
NeurIPS 2023 papers (5K pages),
and Manuallib product manuals (20K pages).
Queries are generated by prompting GPT-4o with each page image.
Training batches are organized so that all samples within a batch of 128 originate from the same source domain.

\paragraph{ColPali Training Set \cite{faysse2025colpaliefficientdocumentretrieval}.}
The ColPali training set comprises 127K query--page pairs, of which 63\% come from existing academic VQA benchmarks
(DocVQA, ArXiv~QA, TatDQA, Infographic-VQA)
and 37\% are synthetically generated from web-crawled PDF pages across energy, healthcare, government, AI, and business domains using Claude-3 Sonnet.
The dataset is English-only by design, enabling study of zero-shot cross-lingual transfer.
After our deduplication pipeline, 109K pairs remain.

\paragraph{VisRAG-InDomain \cite{yu2025visragvisionbasedretrievalaugmentedgeneration}.}
This dataset provides 123K query--page pairs drawn from six established visual QA benchmarks:
PlotQA (56K, scientific plots),
ArXivQA (26K, academic papers),
Infographic-VQA (18K),
MP-DocVQA (11K),
SlideVQA (8K, presentation slides),
and ChartQA (4K).
Since these benchmarks overlap with ViDoRe evaluation sets, we rely on the official train splits to avoid contamination.
After deduplication, 94K pairs remain.

\paragraph{VDR-Multilingual \cite{cimolai2025vdr}.}
This multilingual dataset contains approximately 500K query--image pairs across five European languages (English, German, French, Spanish, Italian), collected from public internet PDFs.
Queries are synthetically generated using Gemini-1.5-Pro and Qwen2-VL-72B.
Each record also includes 16 hard negative document IDs (unused in our align-only training).
After deduplication, 275K pairs remain with a balanced distribution across languages (52--57K per language).
This is the only multilingual source in our training mixture, providing cross-lingual diversity.
The dataset is released under the Apache~2.0 license.

\section{Per-Dataset Results}
\label{app:per-dataset}

Tables~\ref{tab:v1-detail}--\ref{tab:v3-detail} provide the full per-dataset breakdown for all three ViDoRe benchmark versions, covering all baselines and \ours variants (including \ours-S-Multi with multilingual augmentation) evaluated under identical conditions.


\begin{table*}[t]
\centering
\small
\setlength{\tabcolsep}{2pt}
\caption{Per-dataset \ndcg ($\times$100) on ViDoRe v1. $\dagger$: Qwen3-VL-Embedding-2B (our teacher). Best overall in \textbf{bold}, best among \ours variants \underline{underlined}.}
\label{tab:v1-detail}
\begin{tabular}{lrrrrrrrrrrr}
\toprule
 & \multicolumn{8}{c}{English} & \multicolumn{2}{c}{French} & \\
\cmidrule(lr){2-9} \cmidrule(lr){10-11}
Model & ArXivQA & TatDQA & InfoVQA & DocVQA & Syn-AI & Syn-En. & Syn-Gov. & Syn-Hlt. & Shift. & TabFQ. & Avg \\
\midrule
Tomoro-8B & \textbf{91.3} & \textbf{81.5} & \textbf{94.6} & \textbf{66.2} & 98.9 & 96.2 & \textbf{96.9} & 98.7 & 87.6 & 94.3 & \textbf{90.6} \\
Tomoro-4B & 90.9 & 80.8 & 94.2 & 66.0 & 98.5 & 95.9 & 95.7 & 98.9 & 87.4 & 94.0 & 90.2 \\
ColNomic-7B & 87.9 & 80.5 & 91.9 & 60.9 & \textbf{100.0} & \textbf{96.8} & 95.8 & \textbf{99.1} & \textbf{88.7} & \textbf{96.0} & 89.8 \\
ColPali & 82.7 & 70.3 & 85.4 & 58.9 & 96.7 & 94.2 & 96.0 & 96.9 & 75.2 & 85.9 & 84.2 \\
ColModernVBert & 72.7 & 71.3 & 83.5 & 47.2 & 96.7 & 89.0 & 93.0 & 94.7 & 50.1 & 68.9 & 76.7 \\
\midrule
DSE-Qwen2 & 85.6 & 67.5 & 87.2 & 56.3 & 96.6 & 92.3 & 96.0 & 96.5 & 80.3 & 93.0 & 85.1 \\
Teacher$^\dagger$ & 81.8 & 63.0 & 88.4 & 49.3 & 97.4 & 90.9 & 95.3 & 97.0 & 83.4 & \textbf{96.3} & 84.3 \\
\midrule
JinaCLIP & 68.2 & 31.8 & 60.9 & 27.6 & 67.7 & 64.6 & 67.5 & 68.6 & 33.6 & 46.9 & 53.7 \\
SigLIP2 & 35.0 & 18.8 & 56.2 & 22.0 & 55.5 & 57.3 & 42.0 & 62.6 & 29.1 & 67.5 & 44.6 \\
BiModernVBert & 40.4 & 27.5 & 43.0 & 15.2 & 45.2 & 45.2 & 41.2 & 49.5 & 18.7 & 48.0 & 37.4 \\
\midrule
\ours-L & \underline{79.1} & \underline{59.6} & \underline{87.4} & 46.2 & 96.1 & 89.3 & 93.9 & \underline{96.4} & 79.9 & 95.8 & \underline{82.4} \\
\ours-M & 78.2 & 59.5 & 86.0 & \underline{46.6} & 95.5 & 89.6 & 94.3 & 94.9 & 80.8 & \underline{96.0} & 82.1 \\
\ours-S & 77.6 & 58.7 & 86.6 & 45.8 & 95.3 & 90.3 & \underline{95.0} & 95.4 & \underline{81.4} & \underline{96.0} & 82.2 \\
\ours-S-Multi & 78.7 & 58.4 & 85.9 & 46.3 & \underline{96.5} & \underline{91.0} & 93.7 & 94.9 & 80.9 & 95.6 & 82.2 \\
\bottomrule
\end{tabular}
\end{table*}


\begin{table*}[t]
\centering
\small
\setlength{\tabcolsep}{5pt}
\caption{Per-dataset \ndcg ($\times$100) on ViDoRe v2. $\dagger$: Qwen3-VL-Embedding-2B (our teacher). Best overall in \textbf{bold}, best among \ours variants \underline{underlined}.}
\label{tab:v2-detail}
\begin{tabular}{lrrrrr}
\toprule
Model & ESG & BioMed & Econ & ESG-H & Avg \\
\midrule
Tomoro-8B & 60.4 & 65.4 & 59.5 & 74.7 & 65.0 \\
Tomoro-4B & 62.6 & 65.6 & 56.0 & \textbf{76.8} & \textbf{65.2} \\
ColNomic-7B & 53.6 & 63.2 & 55.6 & 69.4 & 60.4 \\
ColPali & 54.9 & 55.9 & 49.4 & 58.7 & 54.7 \\
ColModernVBert & 30.5 & 26.6 & 27.2 & 49.3 & 33.4 \\
\midrule
DSE-Qwen2 & 54.9 & 55.9 & 52.0 & 60.0 & 55.7 \\
Teacher$^\dagger$ & \textbf{65.4} & \textbf{68.4} & \textbf{63.4} & 63.7 & \textbf{65.3} \\
\midrule
JinaCLIP & 16.1 & 31.3 & 35.4 & 24.1 & 26.7 \\
SigLIP2 & 9.4 & 33.4 & 28.3 & 9.4 & 20.1 \\
BiModernVBert & 6.1 & 7.6 & 15.8 & 14.0 & 10.9 \\
\midrule
\ours-L & \underline{63.5} & 62.0 & 59.3 & 61.2 & 61.5 \\
\ours-M & 62.7 & \underline{63.6} & 59.8 & \underline{62.8} & \underline{62.2} \\
\ours-S & 61.1 & 61.0 & 57.8 & 62.1 & 60.5 \\
\ours-S-Multi & 62.0 & 62.7 & \underline{60.5} & 62.5 & 61.9 \\
\bottomrule
\end{tabular}
\end{table*}


\begin{table*}[t]
\centering
\small
\setlength{\tabcolsep}{3.5pt}
\caption{Per-dataset \ndcg ($\times$100) on ViDoRe v3. $\dagger$: Qwen3-VL-Embedding-2B (our teacher). Best overall in \textbf{bold}, best among \ours variants \underline{underlined}.}
\label{tab:v3-detail}
\begin{tabular}{lrrrrrrrrr}
\toprule
Model & Fin-EN & Fin-FR & CS & HR & Energy & Indust. & Pharma & Physics & Avg \\
\midrule
Tomoro-8B & \textbf{62.8} & \textbf{45.9} & 72.4 & \textbf{61.5} & \textbf{65.8} & 52.5 & \textbf{64.3} & \textbf{46.7} & \textbf{59.0} \\
Tomoro-4B & 61.3 & 42.9 & 72.5 & 57.3 & 63.3 & \textbf{52.7} & 64.0 & 46.6 & 57.6 \\
ColNomic-7B & 55.2 & 42.8 & \textbf{74.1} & 57.0 & 62.4 & 49.4 & 61.1 & 45.0 & 55.9 \\
ColPali & 34.9 & 21.8 & 63.6 & 43.4 & 45.7 & 34.6 & 52.7 & 39.5 & 42.0 \\
ColModernVBert & 18.3 & 11.6 & 25.9 & 12.9 & 22.9 & 11.4 & 19.4 & 17.0 & 17.4 \\
\midrule
DSE-Qwen2 & 40.1 & 22.4 & 60.8 & 40.8 & 45.0 & 30.4 & 51.8 & 39.0 & 41.3 \\
Teacher$^\dagger$ & 53.5 & 32.5 & 69.3 & 48.9 & 54.8 & 38.2 & 60.2 & 42.3 & 50.0 \\
\midrule
JinaCLIP & 13.4 & 9.4 & 37.9 & 21.7 & 19.2 & 10.6 & 29.5 & 23.8 & 20.7 \\
SigLIP2 & 6.4 & 4.8 & 28.2 & 13.6 & 14.8 & 12.8 & 20.7 & 17.0 & 14.8 \\
BiModernVBert & 4.5 & 2.0 & 11.6 & 4.3 & 4.8 & 3.7 & 8.2 & 5.1 & 5.5 \\
\midrule
\ours-L & 46.5 & 29.2 & 59.1 & 43.2 & 51.7 & 28.7 & 55.1 & 40.1 & 44.2 \\
\ours-M & 47.0 & 30.2 & 60.3 & 43.1 & 51.8 & 30.2 & 55.1 & 39.9 & 44.7 \\
\ours-S & 45.6 & 30.2 & 57.8 & 42.1 & 50.9 & 27.9 & 54.0 & 39.3 & 43.5 \\
\ours-S-Multi & \underline{48.9} & \underline{32.2} & \underline{62.5} & \underline{45.3} & \underline{52.7} & \underline{32.2} & \underline{57.4} & \underline{40.5} & \underline{46.5} \\
\bottomrule
\end{tabular}
\end{table*}

\section{Per-Backbone Loss Ablation}
\label{app:loss-backbone}

Table~\ref{tab:loss-detail} shows loss ablation broken down by student backbone.

\begin{table*}[t]
\centering
\small
\setlength{\tabcolsep}{4pt}
\caption{Loss ablation (\ndcg $\times$100) per variant. S = \ours-S, M = \ours-M, L = \ours-L.}
\label{tab:loss-detail}
\begin{tabular}{lcc*{3}{r}*{3}{r}*{3}{r}}
\toprule
 & & & \multicolumn{3}{c}{ViDoRe v1} & \multicolumn{3}{c}{ViDoRe v2} & \multicolumn{3}{c}{ViDoRe v3} \\
\cmidrule(lr){4-6} \cmidrule(lr){7-9} \cmidrule(lr){10-12}
Config & $\lambda_a$ & $\lambda_r$ & S & M & L & S & M & L & S & M & L \\
\midrule
Align     & 1   & 0   & 82.2 & 82.1 & 82.4 & 60.5 & 62.2 & 61.5 & 43.5 & 44.7 & 44.2 \\
Align-dom & 1   & 0.5 & 81.4 & 81.6 & 81.8 & 59.5 & 59.7 & 60.2 & 41.9 & 43.5 & 43.1 \\
Combined  & 1   & 1   & 81.2 & 81.4 & 82.0 & 58.6 & 59.4 & 59.3 & 41.4 & 42.9 & 43.0 \\
Rank-dom  & 0.5 & 1   & 81.1 & 81.6 & 81.8 & 57.3 & 58.6 & 59.9 & 40.9 & 42.6 & 42.8 \\
Rank      & 0   & 1   & 80.3 & 81.1 & 81.9 & 56.5 & 57.0 & 58.9 & 40.0 & 41.8 & 42.9 \\
\midrule
InfoNCE   & \multicolumn{2}{c}{hard} & 71.4 & 71.5 & 71.7 & 39.2 & 38.9 & 41.4 & 29.8 & 30.3 & 30.0 \\
\bottomrule
\end{tabular}
\end{table*}

\section{Training Cost: Alignment vs.\ Ranking}
\label{app:training-cost}

Table~\ref{tab:training-cost} quantifies this difference.
For our 711K-pair training set, encoding document images with the 2B-parameter VLM teacher costs 24 GPU-hours (8 shards on H200), while text query encoding completes in under 1 GPU-hour---a 30$\times$ speedup.
Alignment-only training eliminates the image encoding step entirely, reducing the total pre-caching cost from ${\sim}$25 to ${\sim}$1 GPU-hour and halving the embedding storage from 5.8\,GB to 2.9\,GB.

\begin{table}[t]
\centering
\small
\setlength{\tabcolsep}{3pt}
\caption{Pre-caching and training cost comparison for align-only ($\lambda_r\!=\!0$) vs.\ ranking-based ($\lambda_r\!>\!0$) distillation on 711K training pairs with a 2B VLM teacher.}
\label{tab:training-cost}
\begin{tabular}{lcc}
\toprule
Requirement & Align-only & With ranking \\
\midrule
Teacher query encoding (text) & \checkmark & \checkmark \\
Teacher doc encoding (images) & --- & \checkmark \\
\midrule
Query encoding time & 0.8 GPU-h & 0.8 GPU-h \\
Doc encoding time & --- & 24 GPU-h \\
\textbf{Total pre-cache cost} & \textbf{0.8 GPU-h} & \textbf{24.8 GPU-h} \\
\midrule
Embedding storage & 2.9 GB & 5.8 GB \\
Temperature tuning ($\tau_t, \tau_s$) & not needed & required \\
Per-step complexity & $O(B)$ & $O(B^2)$ \\
\bottomrule
\end{tabular}
\end{table}

Combined with the accuracy advantage demonstrated in Table~\ref{tab:loss}, alignment-only distillation is dominant in our setting: it is both more accurate and substantially cheaper.

\section{Data Efficiency}
\label{app:data-efficiency}

\begin{figure}[t]
    \centering
    \includegraphics[width=\columnwidth]{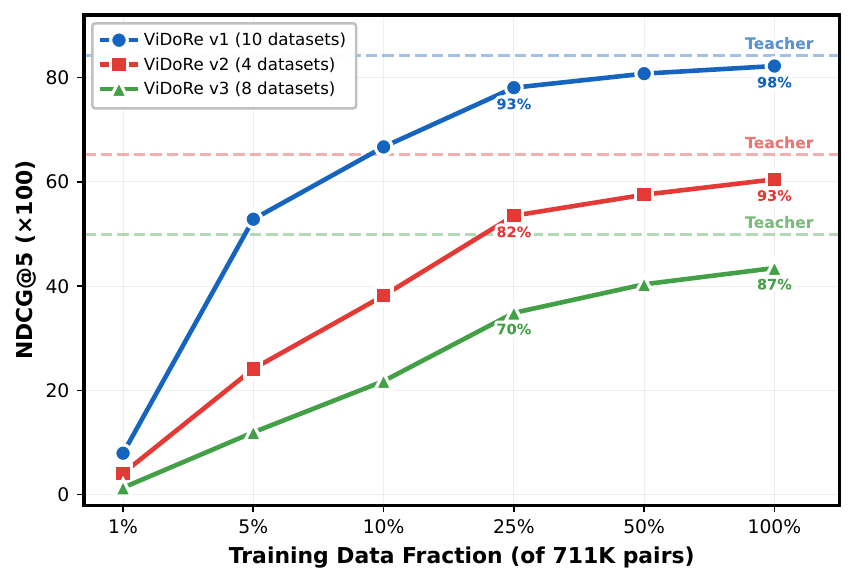}
    \caption{\textbf{Data efficiency of \ours-S.} NDCG@5 vs.\ fraction of 711K training pairs, with teacher upper bounds (dashed). Percentages indicate retention (student/teacher). Diminishing returns are pronounced after 25\% of training data.}
    \label{fig:dataeff}
\end{figure}

We investigate how much training data is needed by training \ours-S (DistilBERT, align-only) on random subsets of \{1, 5, 10, 25, 50, 100\}\% of the 711K training pairs, all with identical hyperparameters (20 epochs, lr=2e-4).
Figure~\ref{fig:dataeff} shows clear diminishing returns: at just 25\% of training data (178K pairs), the model already achieves 93\%/82\%/70\% retention on v1/v2/v3, covering most of the gap to the full-data model (98\%/93\%/87\%).
The marginal gain from quadrupling data (25\%$\to$100\%) adds only +4.2/+7.0/+8.6 NDCG points, compared to +25.2/+29.4/+22.9 from the 5\%$\to$25\% jump.

The three benchmarks saturate at markedly different rates: v1 (English-dominant, smaller corpora) reaches 93\% retention with 25\% data, while v3 (6-language, larger corpora) requires 100\% data to reach 87\%.
This mirrors the cross-lingual finding (\S\ref{sec:cross-lingual}): multilingual evaluation demands greater data diversity.
Even at 10\% (71K pairs), \ours-S achieves 66.7 on v1 (79\% retention), demonstrating that the alignment objective extracts useful cross-modal knowledge from remarkably few examples.

\section{Per-Language Retention Matrix}
\label{app:lang-matrix}

Table~\ref{tab:lang-matrix} presents the full per-language retention breakdown across all 22 ViDoRe datasets.
The 10 ViDoRe v1 datasets are single-language (8~English, 2~French); ViDoRe v2 includes four evaluation languages (EN/FR/ES/DE); ViDoRe v3 includes all six.
Portuguese, entirely absent from training data, is consistently the lowest-retention language across all eight v3 datasets.

\begin{table*}[t]
\centering
\small
\setlength{\tabcolsep}{5pt}
\caption{Per-language retention (\ours-S \ndcg / Teacher \ndcg $\times$ 100) across all 22 ViDoRe datasets. ``---'' = language not present in that dataset. V1 datasets are single-language (8~English, 2~French).}
\label{tab:lang-matrix}
\begin{tabular}{llrrrrrr}
\toprule
Dataset & Ver. & EN & FR & ES & DE & IT & PT \\
\midrule
ArXivQA              & v1 & 94.9 & ---  & ---  & ---  & ---  & --- \\
TatDQA               & v1 & 93.1 & ---  & ---  & ---  & ---  & --- \\
InfoVQA              & v1 & 97.9 & ---  & ---  & ---  & ---  & --- \\
DocVQA               & v1 & 92.8 & ---  & ---  & ---  & ---  & --- \\
SyntheticDocQA-AI    & v1 & 97.8 & ---  & ---  & ---  & ---  & --- \\
SyntheticDocQA-En.   & v1 & 99.3 & ---  & ---  & ---  & ---  & --- \\
SyntheticDocQA-Gov.  & v1 & 99.7 & ---  & ---  & ---  & ---  & --- \\
SyntheticDocQA-Hlt.  & v1 & 98.3 & ---  & ---  & ---  & ---  & --- \\
ShiftProject         & v1 & ---  & 97.6 & ---  & ---  & ---  & --- \\
TabFQuAD             & v1 & ---  & 99.7 & ---  & ---  & ---  & --- \\
\midrule
ESG Reports          & v2 & 95.7 & 96.4 & 93.8 & 87.7 & ---  & --- \\
Biomedical Lectures  & v2 & 89.8 & 90.9 & 91.6 & 84.2 & ---  & --- \\
Economics Reports    & v2 & 95.5 & 91.5 & 95.3 & 82.0 & ---  & --- \\
ESG Reports (Human)  & v2 & 97.4 & ---  & ---  & ---  & ---  & --- \\
\midrule
Finance (EN corpus)  & v3 & 89.7 & 89.9 & 85.5 & 84.5 & 89.0 & 71.3 \\
Finance (FR corpus)  & v3 & 94.0 & 94.2 & 97.9 & 90.6 & 98.1 & 81.8 \\
CS                   & v3 & 92.5 & 84.5 & 83.8 & 82.4 & 84.4 & 72.6 \\
HR                   & v3 & 92.4 & 90.4 & 87.6 & 88.5 & 89.0 & 67.9 \\
Energy               & v3 & 97.1 & 94.9 & 95.1 & 89.8 & 96.3 & 83.3 \\
Industrial           & v3 & 86.5 & 74.3 & 77.9 & 63.3 & 73.5 & 58.7 \\
Pharma               & v3 & 94.9 & 92.3 & 90.1 & 90.7 & 93.4 & 75.9 \\
Physics              & v3 & 95.3 & 93.5 & 95.4 & 90.1 & 92.0 & 91.4 \\
\bottomrule
\end{tabular}
\end{table*}

\section{Residual Factors Beyond Language}
\label{app:residual-factors}

Controlling for language, we analyze English-only queries to identify what corpus-level properties predict the remaining variation in retention. We compute Pearson correlations over 19 datasets with distinct document corpora (Table~\ref{tab:en-corr}), excluding ESG Reports (Human) which shares its corpus with ESG Reports.

\paragraph{Teacher quality is the strongest positive predictor.}
The teacher's own \ndcg correlates most strongly with student retention ($r\!=\!+0.607$): datasets where the teacher achieves higher absolute performance also exhibit higher student retention.
A well-structured teacher embedding space---where relevant documents are clearly separated from irrelevant ones---facilitates distillation, while noisy or ambiguous teacher rankings propagate errors.

\paragraph{Larger corpora are harder to distill.}
The number of documents in the evaluation corpus shows a moderate negative correlation with retention ($r\!=\!-0.566$).
Larger corpora present more confusable candidates, demanding finer-grained query discrimination to preserve top-$k$ rankings.

\paragraph{Student--teacher alignment does not predict retention.}
The mean cosine similarity between student and teacher query embeddings---the direct optimization target of the alignment loss---shows near-zero correlation with dataset-level retention ($r\!=\!+0.094$).
This reveals that retrieval success depends not on how closely the student replicates the teacher's query coordinate, but on whether the relevant document remains among the student's nearest neighbors.
When documents are well-separated (high teacher quality, small corpus), even approximate alignment preserves the correct ranking; when documents cluster densely, even high-fidelity alignment may be insufficient.

\begin{table}[t]
\centering
\small
\setlength{\tabcolsep}{5pt}
\caption{Pearson correlations between corpus-level features and English-only retention (\ours-S, 19 datasets). Teacher = mean teacher \ndcg; $N$ = corpus size; $\bar{c}$ = mean student--teacher query cosine similarity.}
\label{tab:en-corr}
\begin{tabular}{lrrr}
\toprule
 & Teacher & $N$ & $\bar{c}$ \\
\midrule
$r$ with Ret. & +0.607 & $-$0.566 & +0.094 \\
\bottomrule
\end{tabular}
\end{table}

\section{Multilingual Augmentation Pipeline}
\label{app:multilingual-pipeline}

This section details the query-only data augmentation procedure summarized in \S\ref{sec:multilingual-augment}.

\subsection{Translation Procedure}

We first extract all English queries from the four English-language training sources (VisRAG-Synthetic, VisRAG-InDomain, ColPali, VDR-English), yielding 489K source queries.
For each target language, we compute the gap between its current count in the 711K training set and a target of 200K, then randomly sample that many English queries (seed = 42) for translation.
Table~\ref{tab:translation-stats} summarizes the per-language breakdown.

\begin{table}[t]
\centering
\small
\setlength{\tabcolsep}{2pt}
\caption{Translation statistics per target language. Existing = count already in the 711K training set. Translated = newly machine-translated queries. Target = 200K per language.}
\label{tab:translation-stats}
\begin{tabular}{lrrrl}
\toprule
Language & Existing & Translated & Total & Model \\
\midrule
Portuguese &      0 & 200{,}000 & 200{,}000 & en-ROMANCE \\
Italian    & 53{,}787 & 146{,}213 & 200{,}000 & en-it \\
French     & 54{,}079 & 145{,}921 & 200{,}000 & en-fr \\
German     & 56{,}994 & 143{,}006 & 200{,}000 & en-de \\
Spanish    & 57{,}491 & 142{,}509 & 200{,}000 & en-es \\
\midrule
\textbf{Total} & & \textbf{777{,}649} & & \\
\bottomrule
\end{tabular}
\end{table}

Translation uses Helsinki-NLP Opus-MT models \cite{tiedemann-thottingal-2020-opus}, a family of neural machine translation models trained on OPUS parallel corpora.
For Portuguese, we use the multilingual \texttt{opus-mt-en-ROMANCE} model with a \texttt{>>pt<<} target language token; other languages use dedicated bilingual models (\texttt{opus-mt-en-\{es,de,fr,it\}}).
All translations use batch size 64 with maximum sequence length 512.

\subsection{Teacher Embedding Caching}

Each translated query is encoded by the frozen Qwen3-VL-Embedding-2B teacher in text mode, using the same query instruction (\textit{``Find a document image that matches the given query.''}) applied during the original embedding caching.
This produces a 2048-dimensional float16 embedding per translated query.
No document image encoding is involved---the translated queries inherit the original query's positive document pairing, but the alignment-only objective does not use document embeddings during training.
The total teacher encoding cost for 777{,}649 translated queries is under 15 minutes on a single H200 (915 queries/s in text-only mode).

\subsection{Combined Training Set}

The 777{,}649 translated queries are appended to the original 711{,}603 training pairs, yielding 1{,}489{,}252 combined training pairs.
The resulting language distribution is approximately: English 385K (25.9\%), Portuguese 200K (13.4\%), Italian 200K (13.4\%), French 200K (13.4\%), German 200K (13.4\%), Spanish 200K (13.4\%), with the remainder from mixed-language sources.

\subsection{Training Adjustments}

To keep total training steps comparable despite the doubled dataset, we halve the epoch count (10 vs.\ 20) and slightly increase the learning rate (3e-4 vs.\ 2e-4).
All other variables remain identical to the original \ours-S configuration: DistilBERT backbone, align-only loss ($\lambda_a\!=\!1, \lambda_r\!=\!0$), batch size 256 with gradient accumulation 4, MLP projector ($768 \!\to\! 768 \!\to\! 2048$), OneCycleLR with 3\% warmup, and seed 42.

\section{Recomputing SERVAL's ViDoRe v2 Score}
\label{app:serval}

SERVAL \cite{nguyen-etal-2025-serval} reports an average of 63.4 \ndcg under the name ViDoRe v2, but averages nine tasks rather than the four that constitute the benchmark.
The ViDoRe v2 release \cite{macé2025vidorebenchmarkv2raising} comprises four multilingual datasets, and it is those four that Table~\ref{tab:main} averages for every other row.
SERVAL's nine additionally include three English-only slices of three of the same corpora, and two datasets built from AXA documents that have since been withdrawn from the ViDoRe collection.
The English slices are not separate data: each shares its corpus with the corresponding multilingual dataset and holds the English quarter of its queries, so averaging both weights those three corpora twice.

To place SERVAL on the same footing as the other rows, we recompute its average over the four ViDoRe v2 datasets from the per-dataset scores published in that paper.
Table~\ref{tab:serval-recompute} gives the correspondence and the arithmetic.
The result is 62.2 rather than 63.4 for the 32B captioner, and 62.1 for the 72B one; we report the smaller configuration, since the two reach the same nine-task average.

\begin{table*}[h]
\centering
\small
\setlength{\tabcolsep}{4pt}
\begin{tabular}{llrrr}
\toprule
& & \multicolumn{2}{c}{SERVAL reports} & This paper \\
\cmidrule(lr){3-4} \cmidrule(lr){5-5}
ViDoRe v2 & in \cite{nguyen-etal-2025-serval} & 32B+7B & ColNomic & ColNomic \\
\midrule
ESG      & SRSM   & 59.5 & 57.3 & 53.6 \\
BioMed   & SMBTIM & 63.0 & 64.2 & 63.2 \\
Econ     & SEMEM  & 55.5 & 56.7 & 55.6 \\
ESG-H    & RERB   & 70.7 & 73.9 & 69.4 \\
\midrule
\textbf{Average} & & \textbf{62.2} & \textbf{63.0} & \textbf{60.4} \\
\midrule
\multicolumn{5}{l}{\textit{Not part of ViDoRe v2, excluded from the average above}} \\
English slice & SRS   & 59.0 & 54.7 & --- \\
English slice & SMBTI & 65.1 & 66.1 & --- \\
English slice & SEME  & 59.0 & 61.6 & --- \\
withdrawn     & SAXA  & 69.7 & 68.3 & --- \\
withdrawn     & SAXAM & 69.4 & 61.3 & --- \\
\midrule
\multicolumn{2}{l}{\textit{SERVAL's nine-task average}} & 63.4 & 62.7 & --- \\
\bottomrule
\end{tabular}
\caption{Recomputing SERVAL's ViDoRe v2 score. Scores are \ndcg $\times$100, taken from Table 2 of \cite{nguyen-etal-2025-serval}; the short names are that paper's own abbreviations, resolved to ViDoRe datasets through the Hugging Face repository redirects. ColNomic-7B is the one model both papers evaluate, and its two columns bound how far the protocols differ.}
\label{tab:serval-recompute}
\end{table*}

The two evaluations are not otherwise aligned.
ColNomic-7B is the one model both papers report on these four datasets under the same metric, and the two sets of scores differ on all four, by 1.0 to 4.5 points and in the same direction (Table~\ref{tab:serval-recompute}).
The published descriptions do not identify the source of the offset; checkpoint revision, image resolution, and query instruction are all possibilities.
SERVAL's scores are reproduced here as published, with no adjustment.

\section{ViDoRe Query Length Statistics}
\label{app:query-length}

All 19{,}609 ViDoRe query strings (v1 + v2 + v3) as they appear in the datasets, tokenized with the DistilBERT WordPiece tokenizer without truncation. Counts include special tokens. Retrieval evaluation elsewhere in this paper scores the 19{,}537 unique queries left after the exact-duplicate removal applied to ViDoRe v1 (72 duplicated query strings); removing duplicates cannot lengthen the maximum, so the bound reported here holds for both sets. \emph{$>$512} is the number of queries exceeding the student's context limit.

\begin{table*}[h]
\centering
\small
\begin{tabular}{lrrrrrrrr}
\toprule
Dataset & $n$ & min & mean & median & p95 & p99 & max & $>$512 \\
\midrule
\multicolumn{8}{l}{\textit{ViDoRe v1}} \\
\quad ArxivQA          &    500 &   9 &  25.3 &   23 &   42 &   58 &  69 & 0 \\
\quad DocVQA           &    500 &   6 &  11.8 &   11 &   18 &   23 &  28 & 0 \\
\quad InfoVQA          &    500 &   6 &  15.7 &   15 &   25 &   32 &  42 & 0 \\
\quad TabFQuAD         &    280 &  12 &  31.5 &   31 &   45 &   52 &  59 & 0 \\
\quad TAT-DQA          &  1,663 &   7 &  16.9 &   16 &   26 &   32 &  48 & 0 \\
\quad Shift-Project    &    100 &  17 &  30.3 &   30 &   41 &   49 &  52 & 0 \\
\quad SynDocQA AI      &    100 &   9 &  17.0 &   17 &   23 &   26 &  33 & 0 \\
\quad SynDocQA Energy  &    100 &  10 &  18.2 &   17 &   27 &   35 &  36 & 0 \\
\quad SynDocQA Gov     &    100 &   9 &  17.2 &   16 &   23 &   27 &  29 & 0 \\
\quad SynDocQA Health  &    100 &   9 &  17.8 &   17 &   26 &   30 &  32 & 0 \\
\quad \textit{pooled}  &  3,943 &   6 &  18.6 &   17 &   35 &   45 &  69 & 0 \\
\midrule
\multicolumn{8}{l}{\textit{ViDoRe v2}} \\
\quad ESG              &    228 &  12 &  31.3 &   31 &   47 &   53 &  55 & 0 \\
\quad BioMed           &    640 &  13 &  32.2 &   32 &   48 &   54 &  59 & 0 \\
\quad Econ             &    232 &  13 &  30.0 &   31 &   46 &   53 &  56 & 0 \\
\quad ESG-H            &     52 &   4 &  13.1 &   13 &   18 &   23 &  27 & 0 \\
\quad \textit{pooled}  &  1,152 &   4 &  30.7 &   31 &   48 &   53 &  59 & 0 \\
\midrule
\multicolumn{8}{l}{\textit{ViDoRe v3}} \\
\quad Finance EN       &  1,854 &   8 &  35.8 &   34 &   62 &   87 & 111 & 0 \\
\quad Finance FR       &  1,920 &   9 &  35.2 &   33 &   61 &   78 &  97 & 0 \\
\quad CS               &  1,290 &   6 &  37.1 &   32 &   70 &   91 & 182 & 0 \\
\quad HR               &  1,908 &   8 &  40.5 &   38 &   71 &   84 &  98 & 0 \\
\quad Energy           &  1,848 &   8 &  36.2 &   35 &   59 &   70 &  85 & 0 \\
\quad Industrial       &  1,698 &   8 &  41.2 &   40 &   70 &   82 & 102 & 0 \\
\quad Pharma           &  2,184 &   8 &  39.7 &   38 &   68 &   81 & 105 & 0 \\
\quad Physics          &  1,812 &   7 &  38.1 &   36 &   67 &   78 &  90 & 0 \\
\quad \textit{pooled}  & 14,514 &   6 &  38.0 &   35 &   67 &   82 & 182 & 0 \\
\midrule
\textbf{All 22 pooled} & \textbf{19,609} & 4 & 33.7 & 31 & 64 & 78 & 182 & \textbf{0} \\
\bottomrule
\end{tabular}
\caption{DistilBERT-tokenized query lengths across the 22 ViDoRe datasets. The longest single query in the entire benchmark (ViDoRe v3 Computer Science, 182 tokens) uses 35.5\% of the 512-token budget; 99\% of queries fit within 78 tokens (15.2\%).}
\label{tab:query-length}
\end{table*}

\end{document}